%% file: EXO-16-052_temp.tex
\pdfoutput=1

\documentclass[11pt,twoside,a4paper,cmspaper,final,collab]{cms-tdr}
\def\svnVersion{455002P}\def\cmsCernNoTag{CERN-EP-2017-259}\def\cmsCernDate{\today}\def\cmsMessage{Published in the European Physical Journal C as \href{http://dx.doi.org/10.1140/epjc/s10052-018-5740-1}{\doi{10.1140/epjc/s10052-018-5740-1.}}}

\begin{document}\cmsNoteHeader{EXO-16-052}

\hyphenation{had-ron-i-za-tion}
\hyphenation{cal-or-i-me-ter}
\hyphenation{de-vices}
\RCS$HeadURL: svn+ssh://svn.cern.ch/reps/tdr2/papers/EXO-16-052/trunk/EXO-16-052.tex $
\RCS$Id: EXO-16-052.tex 454959 2018-04-11 14:18:26Z ncsmith $

\newcommand{\Hi}{\PH}
\newcommand{\V}{\ensuremath{\mathrm{V}}}
\newcommand{\W}{\PW}
\newcommand{\WW}{\ensuremath{\PW\PW}}
\newcommand{\ZZ}{\ensuremath{\Z\Z}}
\newcommand{\WZ}{\ensuremath{\PW\Z}}
\newcommand{\Lep}{\ensuremath{\ell}}
\newcommand{\ZH}{\ensuremath{\Z\PH}}
\newcommand{\ZHinv}{\ensuremath{\PZ\PH(\text{inv.})}}
\newcommand{\mHi}{\ensuremath{m_{\PH}}}
\newcommand{\mll}{\ensuremath{m_{\Lep\Lep}}}
\newcommand{\met}{\ensuremath{p_{\mathrm{T}}^{\mathrm{miss}}}\xspace}
\newcommand{\dyll}{\ensuremath{\PZ/\gamma^*\to\ell\ell}}
\newcommand{\tw}{\ensuremath{\PQt\PW}}
\newcommand{\usedLumiWithSyst}{\ensuremath{35.9 \pm 0.9\fbinv}}
\newcommand{\LU}{\ensuremath{\Lambda_{\textsf{U}}}\xspace}
\newcommand{\dU}{\ensuremath{d_{\textsf{U}}}\xspace}

\newlength\cmsFigWidth
\newlength\cmsSingleFigWidth
\newlength\cmsDoubleFigWidth
\ifthenelse{\boolean{cms@external}}{\setlength\cmsFigWidth{0.85\columnwidth}}{\setlength\cmsFigWidth{0.4\textwidth}}
\ifthenelse{\boolean{cms@external}}{\setlength\cmsSingleFigWidth{0.40\textwidth}}{\setlength\cmsSingleFigWidth{0.60\textwidth}}
\ifthenelse{\boolean{cms@external}}{\setlength\cmsDoubleFigWidth{0.35\textwidth}}{\setlength\cmsDoubleFigWidth{0.45\textwidth}}
\ifthenelse{\boolean{cms@external}}{\providecommand{\cmsLeft}{top\xspace}}{\providecommand{\cmsLeft}{left\xspace}}
\ifthenelse{\boolean{cms@external}}{\providecommand{\cmsRight}{bottom\xspace}}{\providecommand{\cmsRight}{right\xspace}}
\providecommand{\NA}{{\ensuremath{\text{---}}}}
\newcommand{\x}{\ensuremath{\phantom{0}}}
\newcommand{\y}{\ensuremath{\phantom{.}}}
\ifthenelse{\boolean{cms@external}}
{\providecommand{\suppMaterial}{the supplemental material [URL will be inserted by publisher]}\xspace}
{\providecommand{\suppMaterial}{Appendix~\ref{app:suppMat}\xspace}}

\cmsNoteHeader{EXO-16-052}
\title{Search for new physics in events with a leptonically decaying Z boson and a large transverse momentum imbalance in proton-proton collisions at $\sqrt{s} $ = 13\TeV}
\titlerunning{Search for new physics in Z plus missing transverse momentum events at 13\TeV}

\date{\today}

\abstract{
A search for new physics in events with a Z boson produced in association with large missing transverse momentum at the LHC is presented.
The search is based on the 2016 data sample of proton-proton collisions recorded with the CMS experiment at $\sqrt{s} = 13\TeV$,
corresponding to an integrated luminosity of 35.9\fbinv.
The results of this search are interpreted in terms of a simplified model of dark matter production via spin-0 or spin-1 mediators,
a scenario with a standard-model-like Higgs boson produced in association with the Z boson and decaying invisibly,
a model of unparticle production, and a model with large extra spatial dimensions.
No significant deviations from the background expectations are found, and limits are set on relevant model parameters, significantly extending the results previously achieved in this channel.
}

\hypersetup{%
pdfauthor={CMS Collaboration},%
pdftitle={Search for new physics in events with a leptonically decaying Z boson and a large transverse momentum imbalance in proton-proton collisions at sqrt(s) = 13 TeV},%
pdfsubject={CMS},%
pdfkeywords={CMS, Physics, Exotica, dark matter, simplified model, invisible Higgs}}

\maketitle

\section{Introduction}
\label{sec:intro}
In the pursuit of new physics at the CERN LHC, many scenarios have been proposed in which production of
particles that leave no trace in collider detectors
is accompanied also by production of a standard model (SM) particle, which balances the transverse momentum in an event.
The final state considered in this analysis is the production of a pair of leptons ($\ell^{+}\ell^{-}$, where $\ell=\Pe$ or $\Pgm$),
consistent with originating from a $\PZ$ boson, together with large missing transverse momentum ($\ptmiss$).
This final state is well-suited to probe such beyond the SM (BSM) scenarios, as
it has relatively small and precisely known SM backgrounds.

One of the most significant puzzles in modern physics is the nature of dark matter (DM).
In the culmination of over a century of observations, the ``$\Lambda_{\mathrm{CDM}}$'' standard model of cosmology
has established that, in the total cosmic energy budget,
known matter only accounts for about 5\%, DM corresponds to 27\%, and the rest is dark energy~\cite{2013ApJS..208...19H}.
Although several astrophysical observations indicate that DM exists and interacts gravitationally with known matter,
there is no evidence yet for nongravitational interactions between DM and SM particles.
While the nature of DM remains a mystery, there are a number of models that predict a particle physics origin.
If DM particles exist, they can possibly be produced directly from, annihilate into, or scatter off SM particles.
Recent DM searches have exploited various methods including direct~\cite{Cushman:2013zza} and indirect~\cite{Buckley:2013bha} detection.
If DM can be observed in direct detection experiments,
it must have substantial couplings to quarks and/or gluons, and could also be produced at the LHC~\cite{Beltran:2010ww,Goodman:2010yf,Bai:2010hh,Goodman:2010ku,Fox:2011pm,Rajaraman:2011wf}.

\begin{figure*}[!hbtp]
\centering
\includegraphics[width=0.40\textwidth]{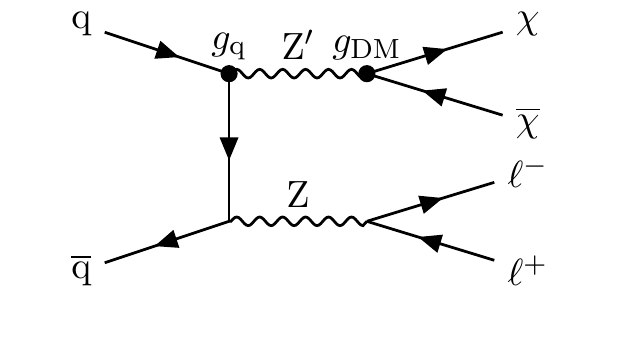} \hfil
\includegraphics[width=0.40\textwidth]{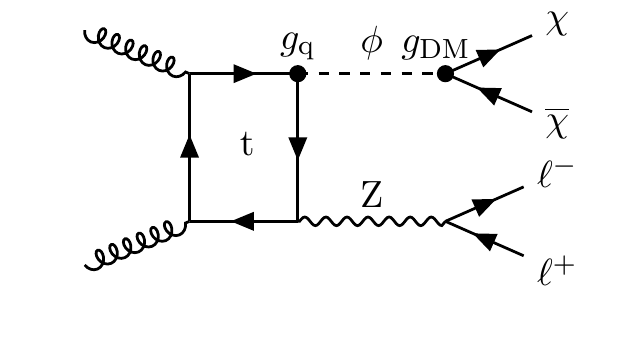} \\
\includegraphics[width=0.40\textwidth]{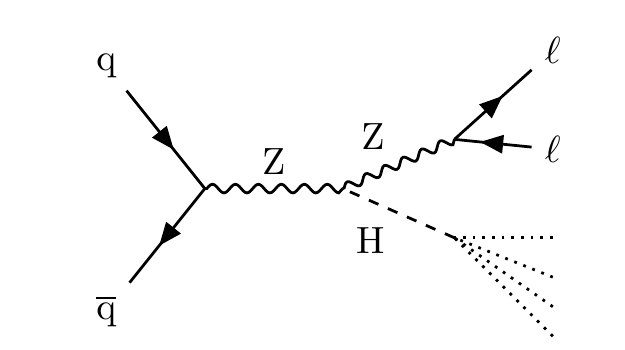} \hfil
\includegraphics[width=0.40\textwidth]{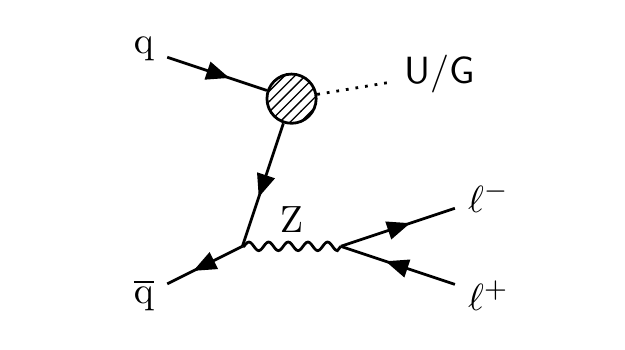}
\caption{Feynman diagrams illustrative of the processes beyond the SM considered in this paper:
(upper left)~DM production in a simplified model with a spin-1 mediator $\PZ'$;
(upper right)~DM production in a simplified model with a spin-0 mediator $\phi$;
(lower left)~production of a Higgs boson in association with Z boson with subsequent decay of the Higgs boson into invisible particles;
(lower right)~unparticle or graviton production. The diagrams were drawn using the {\sc TikZ-Feynman} package~\cite{Ellis:2016jkw}.
}
\label{fig:Feynman}
\end{figure*}

A promising possibility is that DM may take the form of weakly interacting massive particles.
The study presented here considers one possible mechanism for producing such particles at the LHC~\cite{Abercrombie:2015wmb}.
In this scenario, a $\PZ$ boson, produced in proton-proton (pp) collisions, recoils against a pair of DM particles, $\chi\overline\chi$. The $\PZ$ boson subsequently decays into
two charged leptons, producing a low-background dilepton signature, together with $\ptmiss$
due to the undetected DM particles. In this analysis, the DM particle $\chi$ is assumed to be a Dirac fermion.
Four simplified models of DM production via an $s$-channel mediator exchange are considered.
In these models, the mediator has a spin of 1 (0) and vector or axial-vector (scalar or pseudoscalar) couplings to quarks and DM particles.
The free parameters of each model are the masses $m_\text{med}$ and $m_\mathrm{DM}$ of the mediator and DM particle, respectively, as well as the coupling
constant $g_{\Pq}$ ($g_\mathrm{DM}$) between the mediator and the quarks (DM particles).
The vector coupling model can be described with the following Lagrangian:
\begin{equation*}
\mathcal{L}_{\text{vector}} = g_\mathrm{DM} {Z'}_{\mu}\overline{\chi}\gamma^{\mu}\chi  + g_{\Pq} \sum_{\Pq} {Z'}_{\mu} \overline{\Pq}\gamma^{\mu}\Pq,
\end{equation*}
\noindent where the spin-1 mediator is denoted as $\PZ'$ and the SM quark fields are referred to as \PQq and $\overline{\PQq}$.
The Lagrangian for an axial-vector coupling is obtained by making the replacement $\gamma^\mu\to\gamma^5\gamma^\mu$.
In the case of a spin-0 mediator $\phi$, the couplings between mediator and quarks are assumed to be Yukawa-like, with $g_{\Pq}$ acting as a
multiplicative modifier for the SM Yukawa coupling ${y_{\Pq} = \sqrt{2}m_{\Pq}/v}$ (where $v = 246 \GeV$ is the SM Higgs field vacuum expectation value),
leading to the Lagrangian:
\begin{equation*}
\mathcal{L}_{\text{scalar}} = g_\mathrm{DM} {\phi}\overline{\chi}\chi  + g_{\Pq} \frac{\phi}{\sqrt{2}}\sum_{\Pq} y_{\Pq} \overline{\Pq}\Pq.
\end{equation*}
\noindent The Lagrangian with pseudoscalar couplings is obtained by inserting a factor of $i\gamma^5$ into each of the two terms (i.e., $\bar\chi\chi \to i\bar\chi\gamma^5\chi$ and $\bar \Pq \Pq \to i\bar \Pq\gamma^5 \Pq$). Example diagrams of DM production via spin-1 and spin-0 mediators are shown in Fig.~\ref{fig:Feynman} (upper left and right, respectively).

A primary focus of the LHC physics program after the discovery of a Higgs boson (H)~\cite{AtlasPaperCombination,CMSPaperCombination,Chatrchyan:2013lba} by
the ATLAS and CMS Collaborations is the study of the properties of this new particle. The observation of a sizable branching
fraction of the Higgs boson to invisible states~\cite{Ghosh:2012ep,Martin:1999qf,Bai:2011wz} would be a strong sign
of BSM physics.  Supersymmetric (SUSY) models embodying R-parity conservation contain a stable neutral lightest SUSY
particle (LSP), e.g., the lightest neutralino~\cite{Belanger:2001am}, leading to the possibility of decays of the Higgs boson into pairs of LSPs.
Certain models with extra spatial dimensions predict graviscalars that could mix with the
Higgs boson~\cite{Giudice:2000av}.  As a consequence, the Higgs boson could oscillate
to a graviscalar and disappear from the SM brane. The signature would be
equivalent to an invisible decay of the Higgs boson. There could also be contributions
from Higgs boson decays into graviscalars~\cite{Battaglia:2004js}.
With the same effect as the simplified DM models presented earlier, ``Higgs portal'' models~\cite{Baek:2012se,Djouadi:2011aa,Djouadi:2012zc} construct a generic connection between SM and DM particles via a Higgs boson mediator.
This analysis considers decays into invisible particles of an SM-like Higgs boson produced in association with a $\PZ$ boson, as shown in Fig.~\ref{fig:Feynman} (lower left).

{\tolerance = 700
Another popular BSM paradigm considered here is the Arkani-Hamed--Dimopoulos--Dvali (ADD) model with large extra spatial dimensions~\cite{ArkaniHamed:1998rs,ArkaniHamed:1998nn,Han:1998sg}, which
is motivated by the hierarchy problem, i.e., the disparity between the electroweak unification
scale ($M_\mathrm{EW} \sim 1\TeV$) and the Planck scale ($M_\mathrm{Pl} \sim 10^{16}\TeV$).
This model predicts graviton (\cPG) production via the process $\PQq\PAQq \to \PZ + \cPG$. The graviton escapes
detection, leading to a mono-$\PZ$ signature (Fig.~\ref{fig:Feynman}, lower right).
In the ADD model, the apparent Planck scale in four space-time dimensions
is given by $M_\mathrm{Pl}^2 \approx M_\mathrm{D}^{n+2}R^n$, where $M_\mathrm{D}$ is the true Planck scale of
the full $n$+4 dimensional space-time and $R$ is the compactification radius of the extra
dimensions. Assuming $M_\mathrm{D}$ is of the same order as $M_\mathrm{EW}$, the observed large value
of $M_\mathrm{Pl}$ points to an $R$ of order 1 mm to 1 fm for 2 to 7 extra dimensions. The consequence of
the large compactification scale is that the mass spectrum of the Kaluza--Klein graviton states
becomes nearly continuous, resulting in a broad $\PZ$ boson transverse momentum (\PT) spectrum.
\par}

The final BSM model considered in this analysis is the phenomenologically interesting concept of unparticles, which appear in the low-energy limit of conformal field theories.
In the high-energy regime, a new, scale invariant Banks--Zaks field with a nontrivial infrared fixed point is introduced~\cite{Banks:1981nn}.
The interaction between the SM and Banks--Zaks sectors is mediated by particles of large mass scale $M_{\textsf{U}}$, below which the interaction is suppressed and can be treated
via an effective field theory (EFT). The low-energy regime will include unparticles, which have phase space factors equivalent to those of a noninteger
number of ordinary particles~\cite{Kang:2014cia,Rinaldi:2014gha,Cheng:1988zx}. In this analysis, the emission of spin-0 unparticles from SM quarks is considered.
Because of the weakness of the unparticle interactions with the SM fields, the unparticle evades detection.
The EFT Lagrangian used to interpret the results is defined as follows:
\begin{equation*}
\mathcal{L}_{U}  = \frac{\lambda}{\LU^{\dU-1}} \mathcal{O}_{\textsf{U}} \overline{\PQq}\PQq,
\end{equation*}
\noindent where $\lambda$ represents the coupling between the SM and unparticle fields, \LU is the cutoff scale of the EFT, and \dU is the characteristic scaling dimension of the theory.
The unparticle operator is denoted as $\mathcal{O}_{\textsf{U}}$.
A representative Feynman diagram of the interaction is shown in Fig.~\ref{fig:Feynman} (lower right).

The search described in this paper is based on a data set recorded with the CMS detector in 2016,
which corresponds to an integrated luminosity of $\usedLumiWithSyst$ of pp collisions
at a center-of-mass energy of 13\TeV.

The paper is organized as follows: after a brief review of previous work in Section~\ref{sec:review}, followed by a description of the CMS detector in Section~\ref{sec:cms},
we discuss the background composition in Section~\ref{sec:background_overview}. Simulated samples are reviewed in Section~\ref{sec:simulation},
followed by the event reconstruction and event selection description in Sections~\ref{sec:object} and \ref{sec:selection}, respectively.
The details of the background estimation are given in Section~\ref{sec:backgrounds}. The multivariate analysis of invisible Higgs boson decays is summarized in Section~\ref{sec:mva},
followed by the discussion of selection efficiencies and systematic uncertainties in Section~\ref{sec:systematics}. The results are given in Section~\ref{sec:results},
and Section~\ref{sec:summary} summarizes the paper.

\section{Review of previous work}
\label{sec:review}

A search by the CMS Collaboration in the same topology using an earlier data set corresponding to an integrated luminosity
of 2.3\fbinv of pp collisions collected in 2015 at $\sqrt{s}=13\TeV$ found no evidence for BSM physics~\cite{CMS-PAPER-EXO-16-010}.
In addition to the order of magnitude increase in the integrated luminosity, significant differences with respect to the previous analysis include:
new techniques for estimating irreducible backgrounds, which were not viable with the previous data set;
improvements in the event selection; and a broader range of BSM models probed.

In the previous CMS result~\cite{CMS-PAPER-EXO-16-010}, under the same simplified model assumptions as used in this paper, DM mediator masses of up to 290 (300)\GeV were excluded
for fixed vector (axial-vector) couplings of $g_{\Pq}=0.25$ and $g_\mathrm{DM}=1.0$.
Here and in what follows all limits are given at 95\% confidence level (CL), unless explicitly specified otherwise.  Similar DM models have been also probed
in the $\gamma$+\ptmiss~\cite{Aaboud:2017dor} and jet+\ptmiss~\cite{Aaboud:2017buf} topologies at $\sqrt{s} = 13\TeV$ by the ATLAS Collaboration,
excluding mediators with vector (axial-vector) couplings up to masses of 1.2 (1.25) \TeV.
The most stringent limits on DM production in this context were obtained in a CMS analysis of events with a jets+\ptmiss topology performed on
a subset of the present data set, corresponding to an integrated luminosity of 12.9\fbinv~\cite{CMS-PAPER-EXO-16-037}.
In that analysis, mediator masses of up to 1.95\TeV were excluded for both vector and axial-vector couplings.
In the case of a scalar mediator coupled only to quarks and DM particles with $g_{\Pq} = g_\mathrm{DM} = 1$, no exclusion was set.
For the pseudoscalar mediator, under the same assumptions, masses below 430\GeV were excluded.

{\tolerance = 700
Invisible decays of the SM Higgs boson---hereafter H(inv.)---have been
targeted by both ATLAS and CMS.  These searches used both the Z+\ptmiss
and jets+\ptmiss topologies, the latter including gluon fusion and vector
fusion processes as well as associated production with a vector boson
reconstructed as a single jet.
The most stringent constraints were obtained from a combination of searches in these final states at $\sqrt{s} = 8\TeV$ by ATLAS~\cite{Aad:2015pla} and at
multiple center-of-mass energies by CMS~\cite{CMS-PAPER-HIG-16-016}, which, under the assumption of SM production, exclude a branching fraction for H(inv.)
decays larger than $25\%$ for ATLAS and $24\%$ for CMS.
\par}

{\tolerance = 700
Real emission of gravitons in the ADD scenario has been most recently probed in the jet+\ptmiss topology by CMS at $\sqrt{s} = 8\TeV$~\cite{Khachatryan:2014rra}
and by ATLAS at $\sqrt{s} = 13\TeV$~\cite{Aaboud:2016tnv}. In these analyses, the fundamental Planck scale $M_\mathrm{D}$ of the $n$+4 dimensional theory
has been constrained to be larger than 3.3--5.6\TeV (CMS) and 4.1--6.6\TeV (ATLAS), for the number of extra dimensions between $6$ and $2$.
Previous CMS analyses in the same final state as this analysis have excluded unparticle cutoff scales from $400\GeV$ at large values of the scaling dimension
$\dU=2.2$, up to hundreds of \TeV at low values of $\dU\approx1$~\cite{Khachatryan:2015bbl,CMS-PAPER-EXO-16-010}.
\par}

\section{The CMS detector}
\label{sec:cms}

The central feature of the CMS apparatus is a superconducting solenoid of 6\unit{m} internal diameter, providing a magnetic field of 3.8\unit{T}. Within the solenoid volume are a silicon pixel and strip tracker, a lead tungstate crystal electromagnetic calorimeter (ECAL), and a brass and scintillator hadron calorimeter (HCAL), each composed of a barrel and two endcap sections. Forward calorimeters extend the pseudorapidity coverage provided by the barrel and endcap detectors. Muons are detected in gas-ionization chambers embedded in the steel flux-return yoke outside the solenoid.

Events of interest are selected using a two-tiered trigger system~\cite{Khachatryan:2016bia}. The first level, composed of custom hardware processors, uses information from the calorimeters and muon detectors to select events at a rate of around 100\unit{kHz} within a time interval of less than 4\mus. The second level, known as the high-level trigger, consists of a farm of processors running a version of the full event reconstruction software optimized for fast processing, and reduces the event rate to around 1\unit{kHz} before data storage.

A more detailed description of the CMS detector, together with a definition of the coordinate system used and the relevant kinematic variables, can be found in Ref.~\cite{CMSdetector}.

\section{Background composition}
\label{sec:background_overview}

Several SM processes can produce the dilepton+$\ptmiss$ final state.
Since none of the BSM physics signals probed in this analysis are expected to produce a resonance peak in the $\ptmiss$ distribution,
adequate modeling of each SM background process is necessary.
The following SM background processes have been considered in this analysis:
\begin{itemize}
\item[$\centerdot$] $\ZZ \to 2\ell2\nu$ production, which yields the same final state as the signal and contributes
approximately 60\% of the total background.

\item[$\centerdot$] $\WZ \to \ell\nu\ell\ell$ production, where the lepton from the $\W$ boson
decay is not identified either because it fails the lepton identification,
or because it falls outside the detector acceptance or kinematic selections.
This process contributes approximately 25\% of the total background, and the
kinematic distributions are similar to those for the $\ZZ \to 2\ell 2\nu$ process.

\item[$\centerdot$] $\WW \to \ell\nu\ell\nu$ events, where the dilepton
invariant mass falls into the $\Z$ boson mass window.  These events constitute
approximately 5\% of the background.

\item[$\centerdot$] Events with leptonically decaying top quarks (mostly $\ttbar$ and $\tw$), where the dilepton invariant mass falls
into the $\Z$ boson mass window, and which contribute about 5\% of the total background.

\item[$\centerdot$] Drell--Yan (DY) production, $\dyll$, which can produce events with large $\ptmiss$ caused mainly by
jet energy mismeasurement and detector acceptance effects.
It amounts to approximately 5\% of the total background.

\item[$\centerdot$] Triboson processes (e.g., $\PW\PW\PW$), which have a small cross section and contribute less than 1\% of the total background.
\end{itemize}

Processes that were found to have a negligible contribution to the signal region include:
$\PW$+jets, because of the very low probability for a jet to be reconstructed as a lepton and the dilepton system to be within the $\Z$ boson mass window;
the SM process $\Z(\to \ell\ell)\Hi(\to\Z\Z\to4\PGn)$, which is a subset of the $\ZHinv$ signal and accounts for 0.1\% of SM Higgs boson decays;
and ${{\rm gg} \to \Hi(\to \WW)}$, which has similar topology to continuum $\WW$ production but makes a negligible contribution after the full selection.

\section{Simulation}
\label{sec:simulation}

Simulated Monte Carlo (MC) samples are used to estimate backgrounds, to validate the background estimation techniques using control samples in data, to calculate signal efficiency, and to optimize the analysis.

Diboson production (VV, where $\mathrm{V}=$ W or Z) via $\PQq \PAQq$ annihilation, as well as $\ZH$ production
via $\PQq \PAQq$ annihilation and gluon fusion, are generated at next-to-leading order (NLO) in quantum chromodynamics (QCD) with \POWHEG2.0~\cite{Alioli:2008gx,Nason:2004rx,Frixione:2007vw,powheg:2010}.
The ${\rm gg \to \WW}$ and ${\rm gg \to \ZZ}$ processes are simulated at NLO with {\sc mcfm} v7.01~\cite{MCFM}.
The $\PZ$+jets, $\Z\gamma$, $\ttbar$, $\ttbar \V$, and $\mathrm{VVV}$ samples are generated at NLO with either \MGvATNLO v2.3.2~\cite{Alwall:2014hca} or \POWHEG.

{\tolerance = 700
Samples of DM particle production in the simplified model framework are generated using
\textsc{DmSimp}~\cite{Mattelaer:2015haa,Backovic:2015soa,Neubert:2015fka} interfaced with \MGvATNLO v2.4.3.
Samples are generated over a range of values for the masses $m_\text{med}$ and $m_\mathrm{DM}$.
For the vector and axial-vector models, samples are generated at NLO in QCD with up to one additional parton
in the matrix element calculations, and the mediator couplings to the SM and DM fields are set to $g_{\Pq} = 0.25$ and $g_\mathrm{DM} = 1$, respectively.
For the scalar and pseudoscalar models, samples are generated at leading order in QCD, and the couplings are set to $g_{\Pq} = g_\mathrm{DM} = 1$.
This choice of couplings is recommended by the ATLAS/CMS dark matter forum~\cite{Abercrombie:2015wmb} and by the LHC dark matter working group~\cite{Boveia:2016mrp}.
For all DM particle production samples, the central values of the renormalization and factorization scales are set to the $m_\mathrm{T}^{2}$ scale after $k_\mathrm{T}$-clustering of the event.
\par}

Events for the ADD scenario of large extra dimensions and for the unparticle model are generated at leading order (LO) using an EFT implementation
in \PYTHIA 8.205 ~\cite{Sjostrand:2007gs,Ask:2008fh,Ask:2009pv}.
In the ADD case, event samples are produced for $M_\mathrm{D} = 1$, 2, and 3 \TeV, each with $n =$ 2--7.
In order to ensure the validity of the EFT, the signal is truncated for $\hat{s} > M_\mathrm{D}^2$, where $\hat{s}$ is the center-of-mass energy squared of the incoming partons.
Events above this threshold are suppressed by an additional weight of $M_\mathrm{D}^4 / \hat{s}^2$. In general, this procedure has a larger effect for large values of $n$,
for which the distribution of $\hat{s}$ is shifted towards higher values~\cite{Ask:2008fh}.
For the unparticle case, samples are generated for scaling dimensions \dU between $1.01$ and $2.2$, with the cutoff scale \LU set to $15 \TeV$ and the coupling $\lambda$ set to 1.
Since both \LU and $\lambda$ modify the cross sections of the signal prediction, but not its kinematic distributions~\cite{Ask:2009pv}, a simple rescaling of cross sections is
performed to obtain signal predictions for alternative values of these parameters. No truncation is performed for the unparticle signal so that the results can be compared with those
of previous searches.

In all cases, \PYTHIA versions 8.205 or higher is used for parton showering, hadronization, and the
underlying event simulation, using tune CUETP8M1~\cite{GEN-14-001}. The merging of jets from matrix element calculations and parton shower descriptions is done using the MLM~\cite{MLM} (FxFx~\cite{Frederix:2012ps}) scheme for LO (NLO) samples. The NNPDF3.0~\cite{nnpdf} parton distribution function (PDF) set is used, with the order corresponding to the one used
for the signal or background simulation.

For all MC samples, the detector response is simulated using a detailed
description of the CMS detector, based on the $\GEANTfour$
package~\cite{Agostinelli:2002hh}. Minimum bias events are superimposed on the
simulated events to emulate the additional pp interactions per bunch crossing
(pileup). All MC samples are corrected to reproduce the
pileup distribution as measured in the data. The average number of pileup events
per bunch crossing is approximately 23 in the data sample analyzed.

\section{Event reconstruction}
\label{sec:object}

{\tolerance = 700
In this analysis, the particle-flow (PF) event reconstruction algorithm~\cite{CMS-PRF-14-001} is used.
The PF algorithm is designed to leverage information from all CMS detector components to reconstruct
and identify individual particles, namely: electrons, muons, photons, and charged and neutral hadrons.
The reconstructed vertex with the largest value of summed physics-object $\pt^2$ is taken to be the primary $\Pp\Pp$ interaction vertex.
The physics objects are the track-jets, clustered using the jet finding algorithm~\cite{Cacciari:2008gp,Cacciari:2011ma} with the
tracks assigned to the vertex as inputs, and the associated missing transverse momentum, taken as the negative vector sum of the $\pt$ of those jets.
\par}

Electron candidates are reconstructed using an algorithm that combines information
from the ECAL, HCAL, and the tracker~\cite{Khachatryan:2015hwa}.
To reduce the electron misidentification rate, electron candidates are subjected to
additional identification criteria, which are based on the distribution of the
electromagnetic shower in the ECAL, the relative amount of energy deposited in the HCAL in the cluster, a matching of the trajectory of an electron
track with the cluster in the ECAL, and its consistency with originating from the selected primary vertex.
Candidates that are identified as originating from photon conversions in the detector material are removed.

Muon candidate reconstruction is
based on two main algorithms: in the first,
tracks in the silicon tracker are matched to
track stubs (or segments) reconstructed in the muon
detectors; in the second algorithm, a combined fit is performed to signals in
both the silicon tracker and the muon system~\cite{muonid}.
The two resulting collections are merged, with the momentum measurement of the latter
algorithm taking precedence.
To reduce the muon misidentification rate, further identification
criteria are applied on the basis of the number of measurements in the tracker and in
the muon system, the quality of the muon track fit, and its consistency with the
selected primary vertex location.

{\tolerance = 700
Leptons produced in the decay of $\PZ$ bosons are expected to be
isolated from hadronic activity in the event. The isolation is defined
from the sum of the momenta of all PF candidates found in a cone of radius
$R=\sqrt{\smash[b]{(\Delta\eta)^2+(\Delta\phi)^2}} = 0.4$
built around each lepton,
where $\Delta\phi$ and $\Delta\eta$ are, respectively, the
differences in the azimuthal angle (measured in radians) and in the pseudorapidity between the
lepton and the PF candidate. The contribution to the isolation from
the lepton candidate itself is removed.
For muons, the isolation sum is required to be smaller than 15\% of the muon $\pt$.
For electrons in the ECAL barrel (endcap), the limit on this isolation sum is 6.9 (8.2)\% of the electron $\pt$.
In order to mitigate the dependence of the isolation variable on the
number of pileup interactions, charged hadrons are included in the sum
only if they are consistent with originating from the selected primary
vertex of the event.
To correct for the contribution to the isolation sum of neutral
hadrons and photons from pileup interactions,
different strategies are adopted for electrons and muons.
For electrons, a median energy density ($\rho$) is determined on an
event-by-event basis using the method described in Ref.~\cite{FASTJET}.
The contribution of the pileup particles is then estimated as a product of $\rho$ and the effective area of the isolation cone and is subtracted from the isolation sum.
For muon candidates, the correction is performed instead by subtracting
half the sum of the $\pt$ of the charged-hadron candidates in the cone of
interest, which are not associated with the primary vertex.
The factor of one half corresponds to the average ratio of neutral to charged particles in pileup interactions.
\par}

Jets are constructed from PF candidates
using the anti-\kt clustering algorithm \cite{Cacciari:2008gp}
with a distance parameter $R = 0.4$, as implemented in the {\sc fastjet}
package~\cite{Cacciari:2011ma, Cacciari:2006gp}. The jet momentum is defined as
the vectorial sum of all PF candidate momenta assigned to the jet, and is found in
the simulation to be within 5 to 10\% of the true momentum over the entire \pt\
range and detector acceptance used in this analysis. An overall energy subtraction is applied to
correct for the extra energy clustered in jets due to pileup interactions,
following the procedure in Refs.~\cite{FASTJET,Cacciari:2008gn}.
Corrections to the jet energy scale and resolution are derived from measurements both in
simulation and in data of the energy
balance in dijet, multijet, $\gamma$+jet, and leptonic Z+jet events~\cite{Chatrchyan:2011ds,Khachatryan:2016kdb}.

The missing transverse momentum vector, $\ptvecmiss$,  is defined as the projection
of the negative vector sum of the momenta of all reconstructed PF candidates in an event
onto the plane perpendicular to the beams.
Its magnitude is referred to as $\ptmiss$.  Several event-level filters are applied
to discard events with anomalous $\ptmiss$ arising from specific well-understood issues
with the detector components or event reconstruction~\cite{CMS-PAS-JME-16-004}.
Jet energy corrections are propagated to the missing transverse momentum by
adjusting the momentum of the PF candidate constituents of each reconstructed jet.

For the purpose of rejecting events involving top quark production, jets originating from b quark fragmentation (b jets) are identified by ``b tagging.''
The b tagging technique employed is based on the ``combined secondary vertex'' CSVv2 algorithm~\cite{Chatrchyan:2012jua,CMS-PAS-BTV-15-001}.
The algorithm is calibrated to provide, on average, 80\% efficiency for tagging jets originating from b quarks,
and 10\% probability of light-flavor jet misidentification.

For the purpose of rejecting events containing $\tau$ leptons, hadronically decaying $\tau$ leptons ($\tauh$) are identified using the ``hadron-plus-strips"
algorithm~\cite{Khachatryan:2015dfa}. The algorithm identifies a jet as a $\tauh$ candidate if a subset of the particles assigned to the jet
is consistent with the hadronic decay products of a $\tau$ lepton~\cite{Khachatryan:2015dfa}.
In addition, $\tauh$ candidates are required to be isolated from other activity in the event.

\section{Event selection}
\label{sec:selection}
Events with electrons (muons) are collected using
dielectron (dimuon) triggers, with the thresholds of
$\pt > 23$ (17)\GeV and $\pt > 12$ (8)\GeV for the leading and subleading electron (muon), respectively.
Single-electron and single-muon triggers (with $\pt$ thresholds of 27 and 24\GeV, respectively) are also used in order to recover residual trigger inefficiencies.

Events are required to have exactly two ($N_{\ell} = 2$) well-identified, isolated leptons of
the same flavor and opposite electric charge ($\Pep\Pem$ or $\Pgmp\Pgmm$).
The leading electron (muon) of the pair must have $\pt > 25$ (20)\GeV,
while $\pt > 20\GeV$ is required for the subleading lepton.
The dilepton invariant mass is required to be within 15\GeV of the established $\PZ$ boson mass $m_{\PZ}$~\cite{Olive:2016xmw}.
The dilepton $\pt$ ($\pt^{\, \ell\ell}$) must be larger than 60\GeV to reject the bulk
of the $\dyll$ background.
Since little hadronic activity is expected in this final state, events
having more than one jet with $\pt>30~\GeV$ are rejected.
The top quark background is suppressed by applying a b jet veto:
events with at least one b-tagged jet with $\pt > 20\GeV$
reconstructed within the tracker acceptance, $|\eta| < 2.4$, are removed.
To reduce the $\PW\PZ$ background in which both bosons decay
leptonically, events containing additional electrons (muons) with
$\pt > 10$ (5)\GeV and events with loosely identified hadronically decaying
$\tau$ leptons ($\tauh$) with $p_\mathrm{T}>18\GeV$ are removed.

{\tolerance = 700
The event selection is optimized using three variables:
the $\ptmiss$, the azimuthal angle formed between the dilepton \pt and
the missing transverse momentum vector, $\Delta\phi(\ptvec^{\, \ell\ell},\ptvecmiss)$, and the \ptmiss-$\pt^{\, \ell\ell}$
balance ratio, $|\ptmiss-\pt^{\, \ell\ell}|/\pt^{\, \ell\ell}$.
The latter two variables are powerful in suppressing reducible background
processes, such as DY and top quark production.
The selection criteria applied to these variables are
optimized in order to obtain the best expected signal sensitivity for a
wide range of DM parameters that are considered.
For each possible set of selections, the full
analysis is repeated, including the estimation of backgrounds from control samples in data and the systematic uncertainties.
The final selection criteria obtained after optimization are:
$\ptmiss > 100\GeV$, $\Delta\phi(\ptvec^{\, \ell\ell},\ptvecmiss) > 2.6$~rad, and
$|\ptmiss-\pt^{\, \ell\ell}|/\pt^{\, \ell\ell} < 0.4$.
\par}

To avoid positive biases in the $\ptmiss$ calculation due to jet mismeasurement, in events with one jet
a threshold is applied on the azimuthal angle between this jet and the missing transverse momentum,
$\Delta\phi(\ptvec^{\, j},\ptvecmiss) > 0.5\unit{rad}$.
To reduce the contribution from backgrounds such as $\WW$ and $\ttbar$, a requirement on the distance between the
two leptons in the $(\eta,\phi)$ plane, $\Delta R_{\ell\ell} < 1.8$, is applied.

There are two types of analyses performed in this paper.
The main analysis method is based on fitting the \ptmiss spectrum in data after applying the above selection criteria defining the signal region (SR).
For the specific interpretation of this analysis involving invisible decays of the SM (125\GeV) Higgs boson,
a multivariate boosted decision tree (BDT) classifier is employed to increase the sensitivity of the analysis.
We use the following set of twelve variables to train a multiclass BDT classifier:
\begin{itemize}
\item[$\centerdot$]  $\left|\mll-m_{\Z}\right|$ (dilepton mass);
\item[$\centerdot$] $\pt^{\ell 1}$ (leading lepton transverse momentum);
\item[$\centerdot$] $\pt^{\ell 2}$ (subleading lepton transverse momentum);
\item[$\centerdot$] $\pt^{\, \ell\ell}$ (dilepton transverse momentum);
\item[$\centerdot$] $| \eta^{\ell 1} |$ (leading lepton pseudorapidity);
\item[$\centerdot$] $| \eta^{\ell 2} |$ (subleading lepton pseudorapidity);
\item[$\centerdot$] $\ptmiss$       (missing transverse momentum);
\item[$\centerdot$] $m_{T}(\pt^{\ell 1}, $\ptmiss$)$ (leading lepton transverse mass);
\item[$\centerdot$] $m_{T}(\pt^{\ell 2}, $\ptmiss$)$ (subleading lepton transverse mass);
\item[$\centerdot$] $\Delta \phi(\ptvec^{\, \ell\ell},\ptvecmiss)$ (azimuthal separation between dilepton and missing momentum);
\item[$\centerdot$] $\Delta R_{\ell\ell}$ (separation between leptons); and
\item[$\centerdot$] $| \cos \theta^{\rm CS}_{\ell1} |$ (cosine of the polar angle in the Collins--Soper frame~\cite{PhysRevD.16.2219} for the leading lepton).
\end{itemize}

Several classes of event samples are considered for the multiclass BDT: ZH(inv.) signal; ZZ; WZ; DY; and flavor-symmetric or nonresonant backgrounds.
A BDT is trained targeting each class, and the final discriminator is taken to be the likelihood assigned to ZH(inv.) production, normalized to the sum of the likelihoods of all processes.
The SR selection for the BDT analysis is slightly altered from that of the $\ptmiss$-based analysis: the dilepton mass requirement is relaxed to be within $30\GeV$ of the $\PZ$ boson mass,
and the selections on $\Delta \phi(\ptvec^{\, \ell\ell},\ptvecmiss)$, $|\ptmiss-\pt^{\, \ell\ell}|/\pt^{\, \ell\ell}$, and $\Delta R_{\ell\ell}$ are omitted.
The selection for training the BDT additionally requires the missing transverse momentum to be greater than $130\GeV$,
where differentiating between the diboson background and signal is most challenging.
The BDT performance in the untrained region of $100\leq\ptmiss\leq130\GeV$ is found to be adequate, whereas
a BDT trained on event samples including this region was found to have significantly degraded performance in the $\ptmiss>130\GeV$ region.

A summary of the selection criteria for the SR of both the \ptmiss-based analysis and the BDT analysis is given in Table~\ref{tab:selectioncuts}.

\begin{table*}[hbtp!]
\centering
\topcaption{Summary of the kinematic selections for the signal region of both the the $\ptmiss$-based analysis and the BDT analysis.
Where the selections for the two analyses differ, the BDT requirement is given in parentheses.}
{
\ifthenelse{\boolean{cms@external}}{}{\resizebox{\textwidth}{!}}
{
\begin{tabular} {lcc}
\hline
Selection & Requirement  & Reject \\
\hline
$N_{\ell}$                                                     & ${=}2$                                          & $\W\Z$, $\V\V\V$      \\
\multirow{2}{*}{$\pt^{\ell}$}                                  & ${>}25/20\GeV$ for electrons                   & \multirow{2}{*}{QCD} \\
& ${>}20\GeV$ for muons                          &   \\
$\Z$ boson mass requirement                                    & $\left|\mll-m_{\Z}\right| < 15\ (30)\GeV$    & $\W\W$, top quark         \\
Jet counting                                                   & ${\leq}1$ jet with $\pt^{\, j} > 30\GeV$    & $\dyll$, top quark, $\V\V\V$ \\
$\pt^{\, \ell\ell}$                                               & ${>}60\GeV$                                    & $\dyll$           \\
b tagging veto                                                 & CSVv2 $<$ 0.8484                             & Top quark, $\V\V\V$   \\
$\tau$ lepton veto                                             & 0 $\PGt_{\rm h}$ cand. with $\pt^{\tau}>18\GeV$ & $\W\Z$   \\
$\ptmiss$                                                         & ${>}100\GeV$ ($130\GeV$, training only)        & $\dyll$, $\W\W$, top quark   \\
$\Delta \phi(\ptvec^{\, j},\ptvecmiss)$                  & ${>}0.5\unit{rad}$                                  & $\dyll$, $\W\Z$         \\
$\Delta \phi(\ptvec^{\, \ell\ell},\ptvecmiss)$           & ${>}2.6\unit{rad}$ (omitted)                        & $\dyll$           \\
$|\ptmiss-\pt^{\, \ell\ell}|/\pt^{\, \ell\ell}$                         & ${<}0.4$ (omitted)                            & $\dyll$             \\
$\Delta R_{\ell\ell}$                                          & ${<}1.8$ (omitted)                            & $\W\W$, top quark       \\
\hline
\end{tabular}
}
}
\label{tab:selectioncuts}
\end{table*}

\section{Background estimation}
\label{sec:backgrounds}
Background contributions are estimated using combined information from simulation and control regions (CRs) in data.
The normalizations of the dominant background processes are constrained
by using a simultaneous maximum likelihood fit to the SR, as well as to the CRs that are described
in this section. The contributions of minor backgrounds in both SR and CRs are predicted from simulation.

\subsection{Diboson background}

{\tolerance = 700
The $\ZZ$ and $\WZ$ processes contribute to the SR via the
${\ZZ\to\Lep\Lep\nu\nu}$ and ${\WZ\to\Lep\nu\Lep\Lep}$ decay modes, respectively, where the decay products of one boson are not detected.
The background estimate for these processes is improved by selecting CRs with alternative decay modes that not only
provide a normalization based on CRs in data, but also probe the lost-boson \pt distribution, which is expected to be independent of the decay mode.
In this way, the $\ptmiss$ spectra of these processes are constrained with respect to their theoretical predictions.
\par}

The ability of the simulation to correctly model the lost-boson rapidity is important, as the SR rapidity acceptance of the lost
boson is necessarily larger than the rapidity acceptance of the proxy boson in each CR, due to the fact that the visible decay products
of the proxy boson in the CR must be inside the detector acceptance.  The impact of possible data-to-simulation discrepancies in the
high-rapidity portion of diboson background in the SR is suppressed by the fact that, as measured in simulation, the majority of the
$\WZ$ and $\ZZ$ contamination in the SR is comprised of events where the lost boson is within the rapidity range of the CRs.
In addition, the proxy boson rapidity distributions in the CRs (or its visible lepton, in the case of the $\WZ$ CR) show a good agreement between data and simulation.

\subsubsection{The WZ control region}
The $\WZ$ control region is formed from events with three well-reconstructed charged leptons.
In this case, the CR is populated by events with the same decay mode as the SR,
but no leptons are lost to identification or acceptance requirements.
A $\PZ$ boson candidate is selected in the same manner as for the SR, and an additional electron or muon, with identical quality requirements as applied to the leptons in the SR, is required.
To enhance the purity of the $\WZ$ selection, $\ptmiss$ of at least 30\GeV is required,
the invariant mass of three leptons is required to be larger than $100\GeV$,
and the invariant masses of all opposite-sign, same-flavor lepton pairs are required to be larger than $4\GeV$.
Backgrounds in this CR are similar to those in the SR, with a sizeable nonprompt background from the DY+jets process, where a jet is misidentified as a lepton.
All background estimates for this CR are taken from simulation.

The $\PW$ boson $\pt$ (``emulated \ptmiss'') is estimated by calculating the vectorial sum of the $\ptvecmiss$
vector and the transverse momentum vector ($\ptvec$) of the third charged lepton.
In simulation, the majority (over $70\%$) of $\WZ$ background contamination in the signal region
originates from events where over $90\%$ of the $\PW$ boson transverse momentum is carried by one
or more neutrinos from the $\PW$ boson decay.  Thus, the majority of the W boson rapidity distribution in the SR is central, although it is less central than in the WZ CR.
Neither the SR nor the $\WZ$ CR topology can probe the
W boson rapidity directly.  However, for the $\WZ$ CR, good agreement between data and simulation in the third
lepton pseudorapidity distributions is observed.

A minor source of WZ background contamination in the SR originates from events where the visible lepton from a $\PW$ boson decay failed identification requirements.
Data-to-simulation discrepancies in this contribution would also manifest in the measured $\WZ$ CR $\ptmiss$ distribution, for which no such mismodeling effects are evident.

\begin{figure*}[htb]
\centering
\includegraphics[width=\cmsDoubleFigWidth]{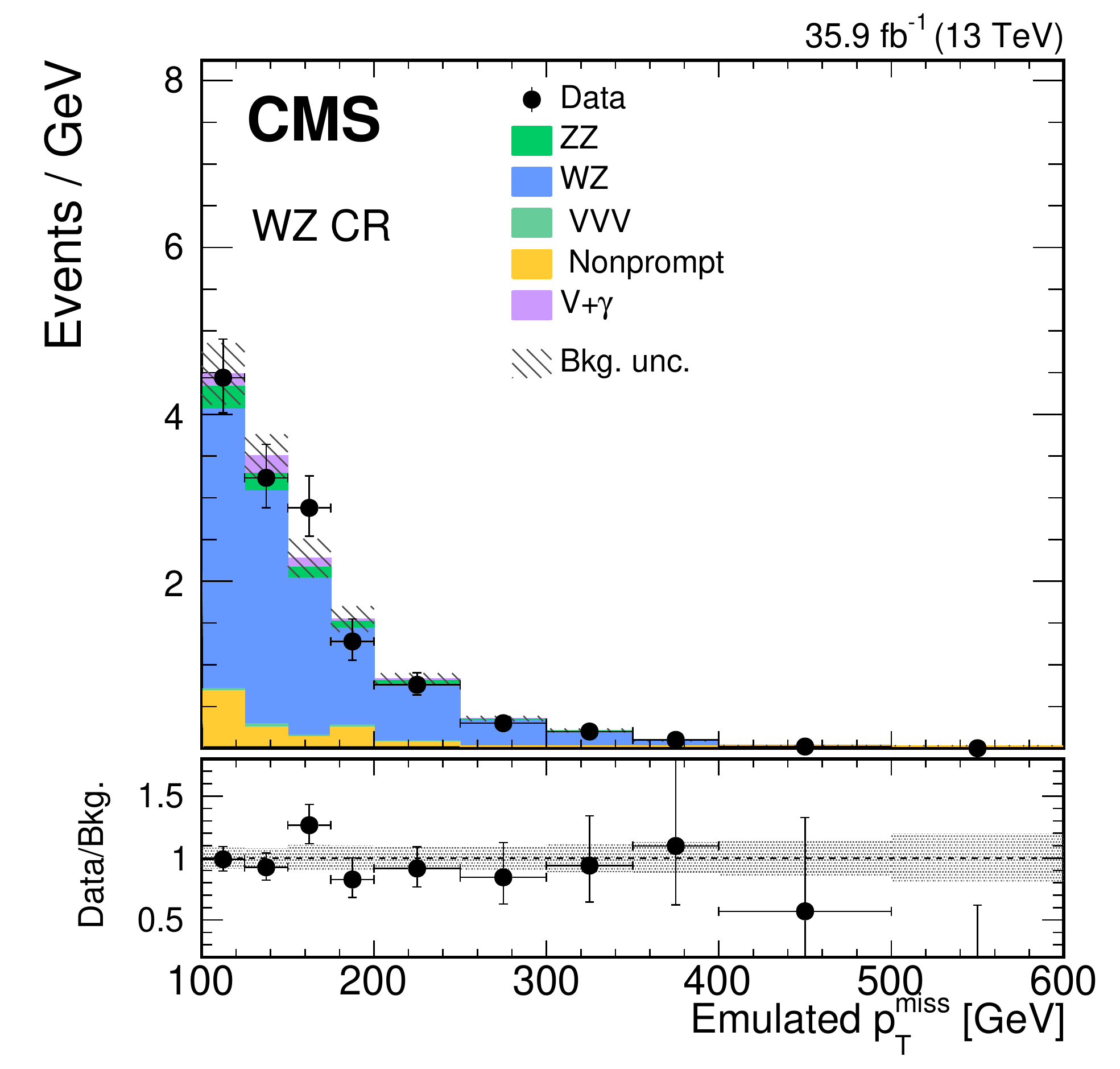} \hfil
\includegraphics[width=\cmsDoubleFigWidth]{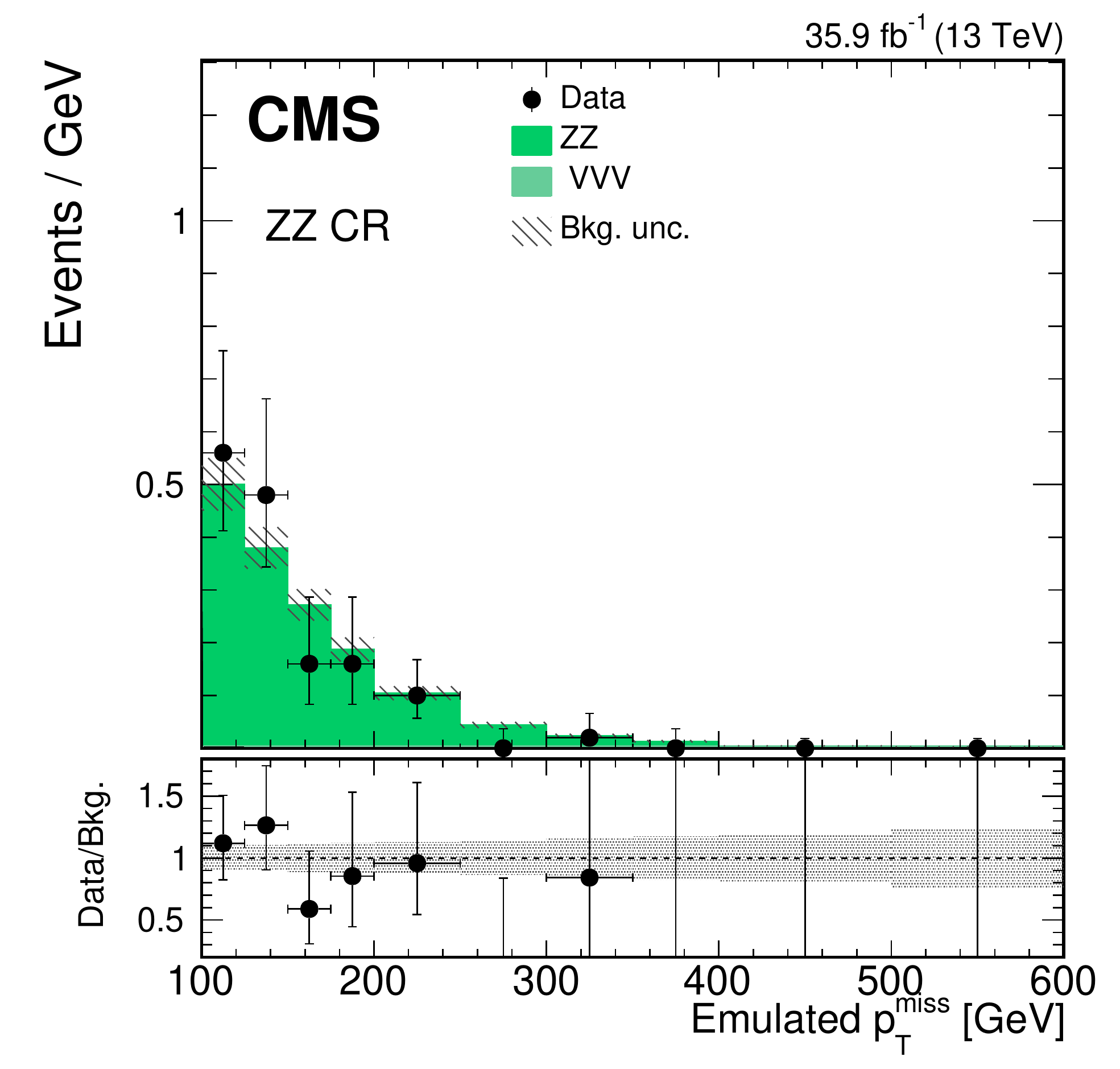} \\
\includegraphics[width=\cmsDoubleFigWidth]{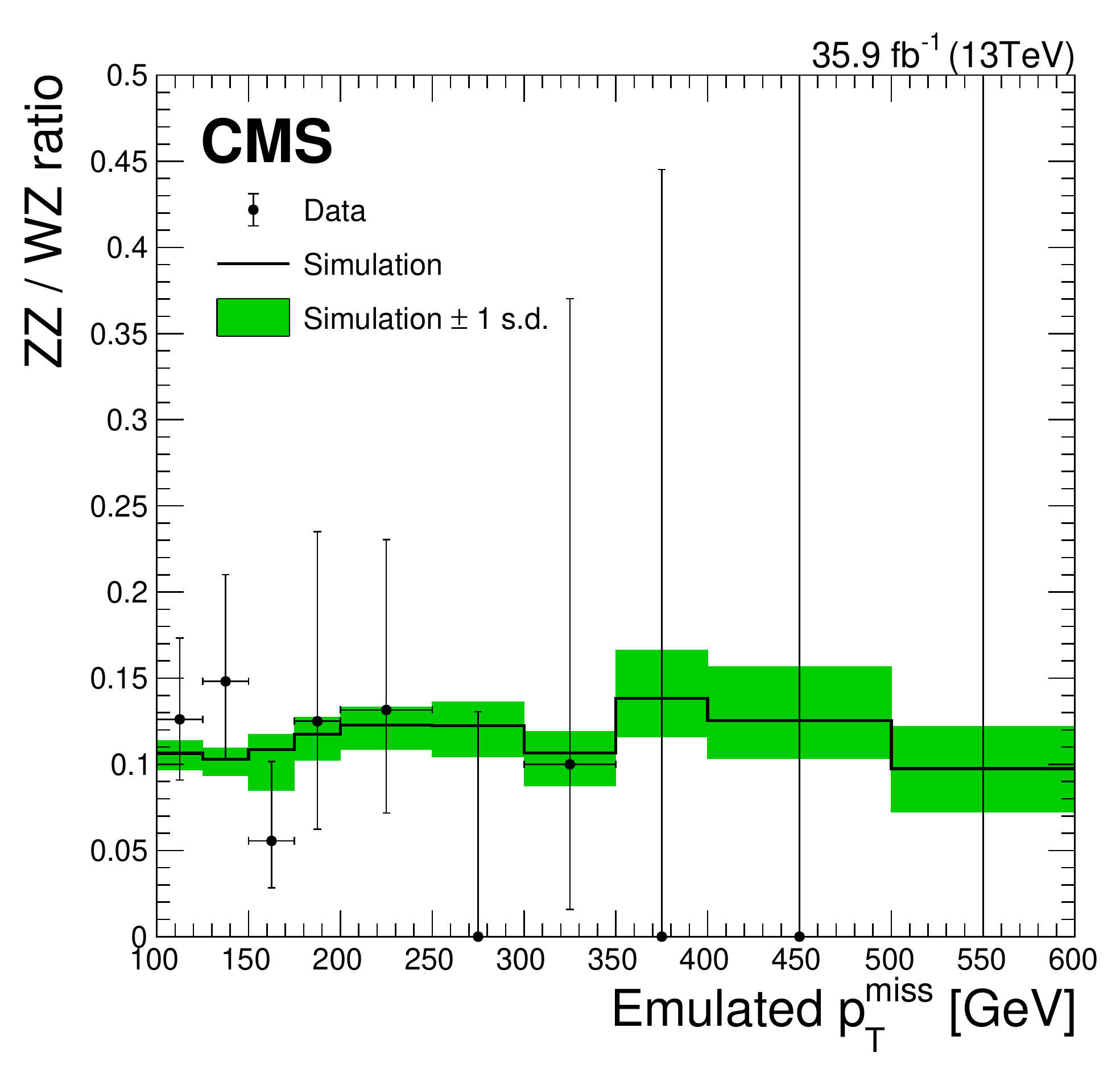}
\caption{
Emulated $\ptmiss$ distribution in data and simulation for the $\PW\PZ \to 3\Lep\nu$ (upper left) and $\Z\Z \to 4\Lep$ (upper right)
CRs, and the ratio between both distributions (lower).
No events are observed with emulated $\ptmiss>500\GeV$ in either channel.
Uncertainty bands correspond to the combined statistical and systematic components.
}
\label{fig:histo_fakemet}
\end{figure*}

Using the emulated $\ptmiss$ in place of the reconstructed $\ptmiss$, the same selection is applied as for the SR.
However, since there is no danger of CR contamination from $\WZ\to\tau\nu\Lep\Lep$ or top quark backgrounds,
no veto on additional $\tauh$ or b jet candidates is applied.
The resulting emulated $\ptmiss$ spectrum is shown in Fig.~\ref{fig:histo_fakemet} (upper left).

\subsubsection{The ZZ control region}
The $\ZZ$ control region is formed from events with four well-reconstructed charged leptons.
In addition to a signal-like $\PZ$ boson candidate, a second $\PZ$ boson candidate is required, the constituents of which only need to pass relaxed lepton quality requirements.
This choice reflects the very high purity of the four-lepton selection. For both candidates, the same $\PZ$ boson mass constraint as in the SR is applied.
Backgrounds, dominated by triboson processes, are almost negligible in this CR and are taken from simulation.

Similar to the $\WZ$ case, the emulated $\ptmiss$ is calculated as the vectorial sum of the $\ptvecmiss$ and the $\ptvec$ of the
$\PZ$ boson with the larger mass difference to the nominal value of $m_{\PZ}$ of the two identified in the event.
The choice of which $\PZ$ boson to use as a proxy for an invisibly decaying one does not significantly alter the emulated $\ptmiss$ spectrum.
In this CR, the rapidity of the proxy boson is observable, for which good agreement between data and simulation is found.

The same selection as in the SR is then applied using the emulated $\ptmiss$ in place of the reconstructed $\ptmiss$, with the exception of the $\tau$ lepton and b jet vetoes.
The resulting emulated $\ptmiss$ spectrum is shown in Fig.~\ref{fig:histo_fakemet} (upper right).

\subsubsection{The VV ratio constraints}
Due to a limited event count in the $\ZZ$ control region, the normalizations of the $\WZ$ and $\ZZ$ processes in the $\WZ$ and $\ZZ$ CRs and the SR are controlled by a single free parameter in the maximum likelihood fit,
with their relative normalizations fixed by the theoretical predictions for the $\WZ$ and $\ZZ$ processes in each \ptmiss bin.
The predictions for these processes are obtained from fully reconstructed simulated events generated as described
in Section~\ref{sec:simulation} with the following additional higher-order corrections applied:
\begin{itemize}
\item[$\centerdot$]
a constant (approximately $10\%$) correction for the $\WZ$ cross section from NLO to NNLO in QCD calculations~\cite{Grazzini:2016swo};
\item[$\centerdot$]
a constant (approximately $3\%$) correction for the $\WZ$ cross section from LO to NLO in electroweak (EW) calculations, considering also photon-quark initial states, according to Ref.~\cite{PhysRevD.88.113005};
\item[$\centerdot$]
a $\Delta\phi(\PZ,\PZ)$-dependent correction, varying in magnitude up to $15\%$, to $\ZZ$ production cross section from NLO to next-to-next-to-leading order (NNLO) in QCD calculations~\cite{Grazzini:2015hta};
\item[$\centerdot$]
a $\pt$-dependent correction, varying in magnitude up to $20\%$ at high $\ptmiss$, to the $\ZZ$ cross section from LO to NLO in EW calculations, following Refs.~\cite{Bierweiler:2013dja,Gieseke:2014gka,PhysRevD.88.113005}, which is the dominant correction in the signal region.
\end{itemize}
We use the product of the above NLO EW corrections and the inclusive NLO QCD corrections~\cite{Campbell:2011bn} as an estimate of the
missing $\textrm{NLO EW}\times\textrm{NLO QCD}$ contribution, which is not used as a correction, but rather assigned as an uncertainty.
The uncertainties in the $\WZ$ and $\ZZ$ EW corrections are assumed to be anticorrelated as a conservative measure.
The uncertainty associated with the NNLO QCD corrections for both processes is represented by the QCD scale variation uncertainties
evaluated on the NLO QCD simulation sample for the respective process, as described in Section~\ref{sec:systematics}.
Figure~\ref{fig:histo_fakemet} (lower) shows the ratio of $\ZZ$ to $\WZ$ CR yields per $\ptmiss$ bin, which probes
the validity of taking the relative normalizations from simulation.
Good agreement is observed between data and simulation.

\subsection{Nonresonant backgrounds}

The contribution of the nonresonant flavor-symmetric backgrounds is estimated from a CR based on events with two leptons of different flavor
($\Pe^{\pm}\Pgm^{\mp}$) that pass all other analysis selections.
Nonresonant background (NRB) consists mainly of leptonic \PW\ boson decays in
$\ttbar$, $\PQt\PW$, and $\PW\PW$ events, where the dilepton mass happens to fall inside the $\PZ$ boson mass window.
Small contributions from single top quark events produced via
$s$- and $t$-channel processes, and $\PZ\to \PGt\PGt$
events in which $\PGt$ leptons decay into light leptons and neutrinos are also
considered in the NRB estimation.

The method assumes lepton flavor symmetry in the final states of these processes.
Since the leptonic decay branching fraction to the $\Pe\Pe$, $\PGm\PGm$, and $\Pe\PGm$ final states from NRB are 1:1:2,
the $\Pe\PGm$ events selected inside the $\PZ$ boson mass window can be extrapolated to the $\Pe\Pe$ and $\PGm\PGm$ channels.
To account for differences in efficiency for electrons and muons, a correction factor $k_{\Pe\Pe}$ is derived
by comparing the NRB yields for the $\Pe\Pe$ and $\PGm\PGm$ channels:
$$k_{\Pe\Pe} = \frac{\epsilon_{\Pe}}{\epsilon_{\PGm}} = \sqrt{\frac{N^{\Pe\Pe}_\mathrm{NRB}}{N^{\PGm\PGm}_\mathrm{NRB}}}$$
under the assumption that there are no efficiency correlations between the two leptons.
In simulation, $k_{\Pe\Pe}$ is found to be about $0.88$ for the final selection.
With this correction factor, the relation between the NRB yields in the SR and CR is:
\begin{equation*}
N^{\ell\ell}_\mathrm{NRB} = \frac{1}{2} \left( k_{\Pe\Pe} + \frac{1}{k_{\Pe\Pe}} \right) N^{\Pe\PGm}_\mathrm{NRB}.
\end{equation*}

The ratio of the NRB contributions in the SR and CR is fixed by this relation. Their
normalization is controlled by a common scaling parameter that is left to float in the maximum likelihood fit.
Perturbations in the predicted transfer factor due to data-to-simulation discrepancies in $k_{\Pe\Pe}$ are suppressed upon summing the $\Pe\Pe + \PGm\PGm$ channels.
The uncertainty in the transfer factor is set conservatively to 20\%.

\subsection{The Drell--Yan background}
The DY background is dominant in the region of low $\ptmiss$.
This process does not produce undetectable particles, therefore any nonzero $\ptmiss$ arises from
the limited detector acceptance and mismeasurement.
The estimation of this background uses simulated DY events, for which the normalization is taken from data in a sideband CR
of $50 \leq \ptmiss \leq 100 \GeV$, with all other selections applied.
In two CRs where a larger DY background contribution is expected, regions with inverted selections
on $\Delta \phi(\ptvec^{\, \ell\ell},\ptvecmiss)$ and on $|\ptmiss-\pt^{\, \ell\ell}|/\pt^{\, \ell\ell}$,
the simulation is found to model the data well.
The sideband CR is included in the maximum likelihood fit, for which the normalization factor is found to be consistent with unity, and
a 100\% uncertainty is assigned to the resulting DY estimate in order to cover the extrapolation from this CR to the SR.
This uncertainty has little effect on the results owing to the small overall contribution from the DY process in the high-$\ptmiss$ SR of this analysis.

\section{Multivariate analysis}
\label{sec:mva}
\begin{figure*}[htbp]
\centering
\includegraphics[width=\cmsDoubleFigWidth]{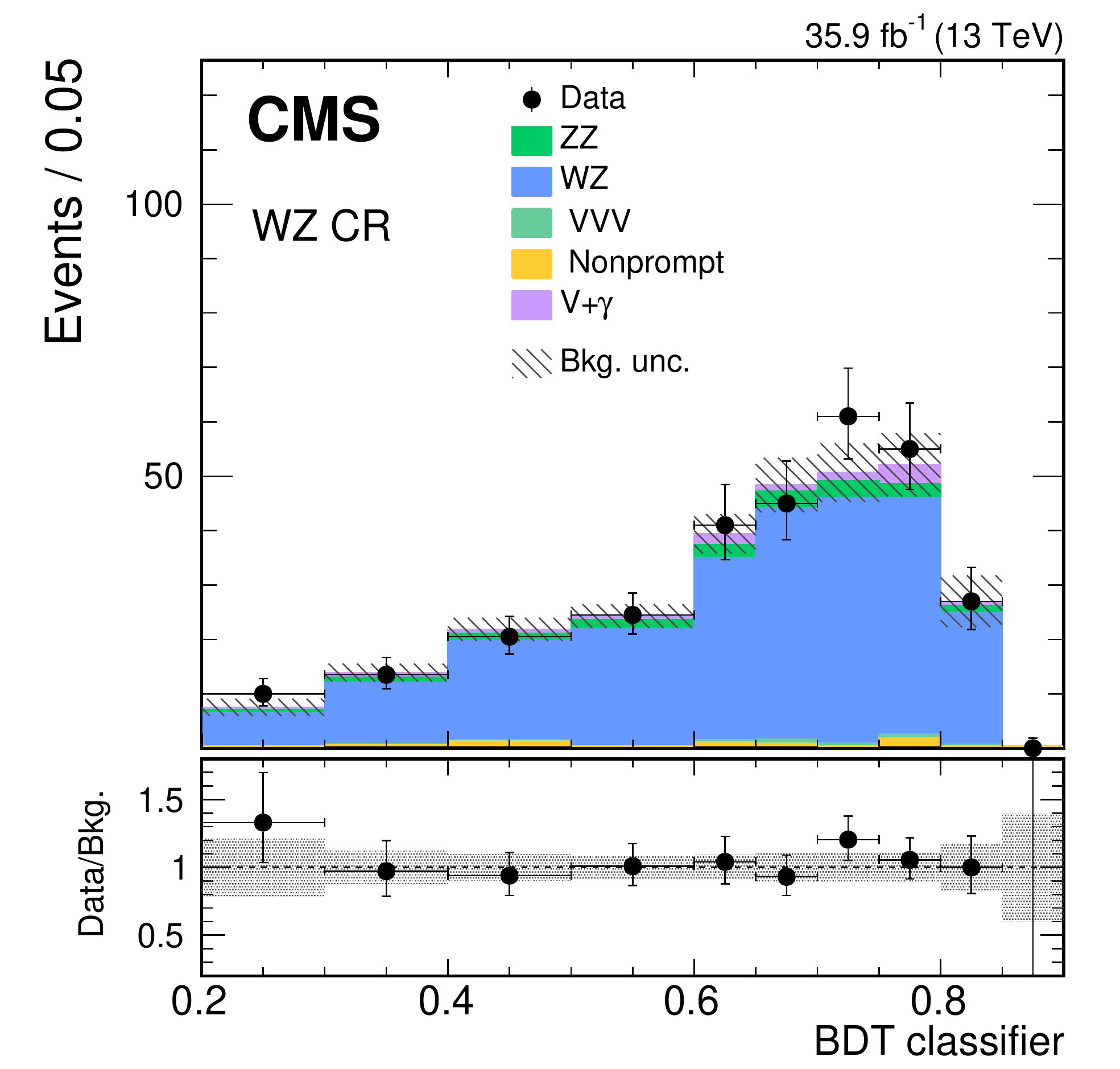} \hfil
\includegraphics[width=\cmsDoubleFigWidth]{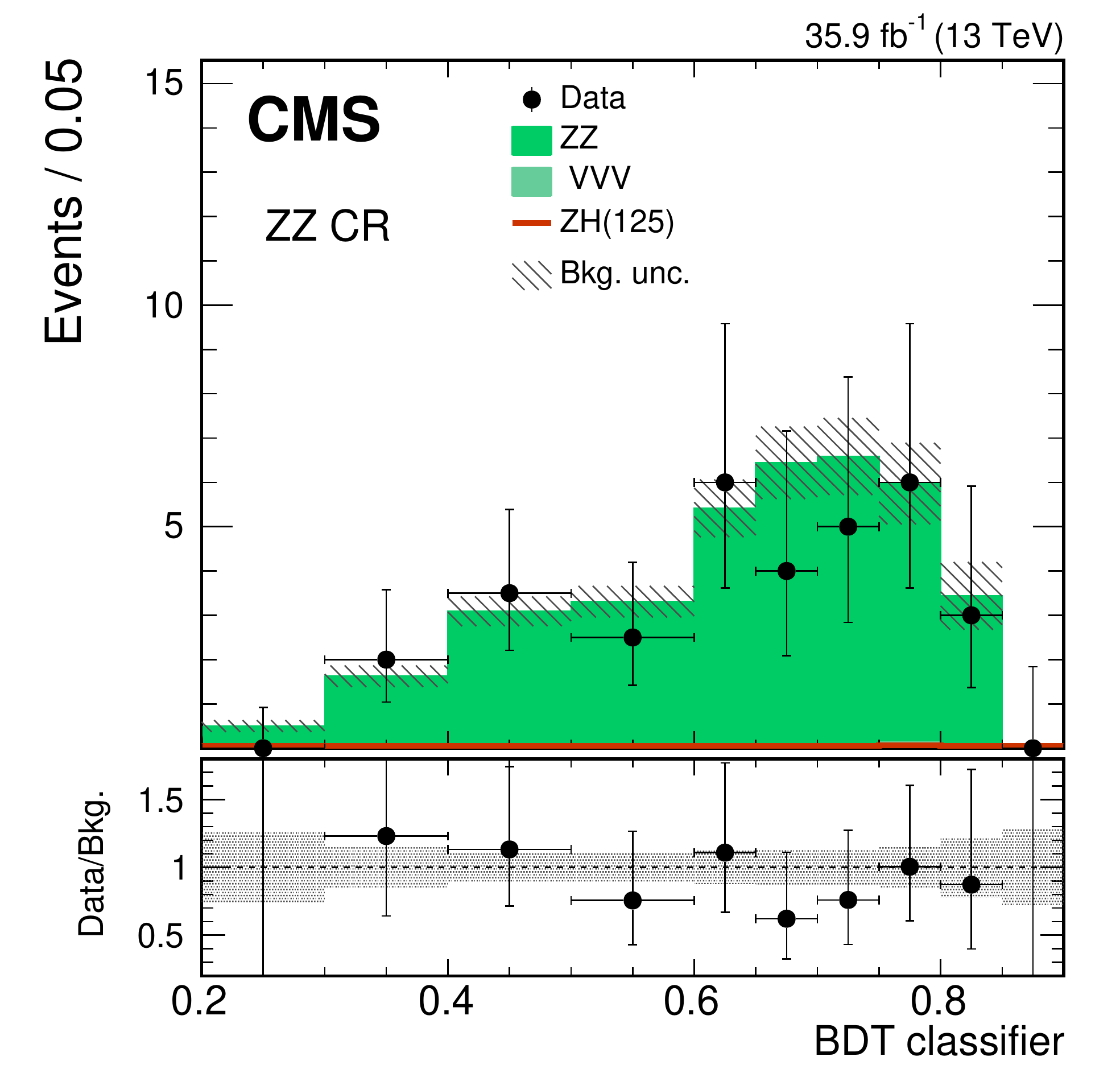}
\caption{Distribution of the BDT classifier in the diboson CRs: (left) $\WZ$ CR; (right) $\ZZ$ CR. Uncertainty bands correspond to the combined statistical and systematic components.}
\label{fig:bdt_vv}

\end{figure*}

For the specific interpretation of this analysis involving invisible decays of the SM (125\GeV) Higgs boson,
a maximum likelihood fit is performed to the spectrum of the BDT classifier values for events satisfying the BDT SR criteria described in Section~\ref{sec:selection}, with the classifier value between 0.2 and 1.
The CR strategy is identical to that in the $\ptmiss$-based analysis, as described in Section~\ref{sec:backgrounds}.
The three- and four-lepton events shown in Fig.~\ref{fig:bdt_vv} are chosen using the same CR selections as in the $\ptmiss$-based analysis.

The multivariate classifier improves the sensitivity of the analysis to the SM H(inv.) model by 10\% compared to the $\ptmiss$-based analysis.
Other than the $\ptmiss$ itself, the variables that provide the most discrimination power are the transverse
masses of each lepton with respect to the $\ptvecmiss$, along with the azimuthal separation between the $\ptvecmiss$ and the dilepton system momentum.
Utilization of this classifier for the other signal models considered in this paper was not pursued, as many of the models' kinematic distributions can vary considerably over the relevant parameter space.

\section{Efficiencies and systematic uncertainties}
\label{sec:systematics}
The efficiency for all backgrounds is estimated using simulation.
The uncertainties in the yields from missing higher-order corrections in signal as well as $\ZZ$ and $\WZ$ background cross sections are evaluated
by independently varying up and down the factorization and renormalization scales by a factor of two.
The effect of these variations on the yields is between 5 and 10\%.
For the $\ZZ$ and $\WZ$ backgrounds, additional uncertainties related to known higher-order
corrections are applied, as discussed in Section~\ref{sec:backgrounds}.

For the Higgs boson signal, the PDF and $\alpha_{s}$ uncertainties comprise the
cross section normalization uncertainties computed by the LHC Higgs cross section working group~\cite{bib:LHCHXSWG2015}
and the effect on the signal acceptance of varying the PDFs and $\alpha_{s}$ within their uncertainties~\cite{Butterworth:2015oua}.
For other signal models, as well as the $\WZ$ and $\ZZ$ backgrounds, the effects of the PDF and $\alpha_{s}$ uncertainties in the signal acceptance are taken into account following the PDF4LHC prescription~\cite{Butterworth:2015oua}.
The PDF and $\alpha_{s}$ uncertainties on these processes are found to be about 1--2\%.

The efficiencies for triggering on, reconstructing, and identifying isolated leptons are obtained
from simulation, and corrected with scale factors determined via a ``tag-and-probe'' technique~\cite{CMS:2011aa} applied to $\Z \to\ell^+\ell^-$ events in data.
The associated uncertainty is about 1--2\% per lepton. An additional
3\% uncertainty associated with the $\WZ \to \ell\PGn\ell\ell$ events, where the reconstructed lepton from the $\PW$ boson decay fails identification, is also included.

In order to reproduce b tagging efficiencies and light-flavor jet mistag rates observed in data, an event-by-event reweighting using data-to-simulation scale factors~\cite{CMS-PAS-BTV-15-001}
is applied to simulated events.
The uncertainty associated with this procedure is obtained by varying the event-by-event weight by ${\pm}1$ standard deviation (s.d.).
The impact on the final yields due to the b tagging efficiency and mistag rate uncertainties is around 1\% for both signal and background.

The impacts of the jet energy scale and resolution uncertainties are
estimated by shifting reconstructed jet energies in simulation by ${\pm}1$ s.d.,
and each is found to have an effect of about 2\% on the yields of the simulated processes after all selections are applied.
The impacts of the electron and muon energy scales are evaluated in the same manner, and have a similar effect.
Uncertainties in the $\ptmiss$ measurement due to the energy resolution of unclustered PF candidates (i.e., those not associated with an electron, muon, or jet) amount to about 2\%.

The uncertainty in the expected yields due to the finite size of the MC samples is considered, and is around 1\% for the signal and main backgrounds.
The simulated MC samples are reweighted to reproduce the pileup conditions observed in data.
The uncertainty related to this procedure is obtained by varying the central value of the estimated inelastic cross section by 5\%~\cite{Aaboud:2016mmw},
and is found to be below 1\%.
The uncertainty assigned to the integrated luminosity measurement is 2.5\%~\cite{CMS-PAS-LUM-17-001}.

The effect of the systematic uncertainties on the shape of the distribution of the discriminating variable ($\ptmiss$ or BDT classifier)
is taken into account by varying the value of the quantity associated with the uncertainty, and observing the resulting variations in the individual bins of $\ptmiss$.

In addition to all of the sources of systematic uncertainty in the $\ptmiss$-based analysis, the following systematic uncertainties in the BDT-based analysis affect the BDT classifier shape.
The most important sources of uncertainty in the BDT classifier shape are the lepton energy scale and \ptmiss uncertainties; their impact on the signal ($\WZ$ and $\ZZ$ backgrounds) amounts
to about 2 (6)\% and translates into an additional 2\% uncertainty in the expected limit on the H(inv.) branching fraction.

All these sources of uncertainty are summarized in Table~\ref{tab:syst}.
The combined uncertainty in the signal efficiency and acceptance is estimated to be about 5\%
and is dominated by the theoretical uncertainty due to missing
higher-order corrections and PDF uncertainties. The total uncertainty in the
background estimations in the signal region is about 15\%,
dominated by the theoretical uncertainties in the $\ZZ$ and $\WZ$ process description.

\begin{table*}[htb]
\renewcommand{\arraystretch}{1.05}
\centering
\topcaption{
Summary of the systematic uncertainties for the \ptmiss- and BDT-based analyses.
Each uncertainty represents the variation of the relative yields of the processes in the SR.
Each uncertainty is fully correlated across processes to which it contributes, including those processes that are also present in CRs. The symbol ``\NA'' indicates that the systematic uncertainty does not contribute or is deemed negligible.
For minor backgrounds, systematic uncertainties are omitted because of the smallness of their contribution.
For shape uncertainties (indicated with a *), the numbers correspond to the overall effect of the shape variation on the yield or acceptance.
The impact on the expected upper limit for the signal strength, i.e., the relative decrease in the median expected upper limit for the signal strength upon removing the nuisance term, is evaluated with respect to the SM H(inv.) signal and presented in the last column. In this column the number in parentheses shows the impact on the BDT-based analysis, if different from that for the \ptmiss-based analysis. The last part of the table provides the additional uncertainties in the BDT-based analysis.
}
{
\begin{tabular}{lcccccc}
\hline
\multirow{2}{*}{Source of uncertainty}         & \multicolumn{5}{c}{Effect (\%)}                                 & Impact on the   \\
& Signal     & \ZZ        & \WZ        & NRB & DY         & exp. limit  (\%)\\
\hline
* VV EW corrections                   & \NA         & 10         & $-4$         & \NA          & \NA         & 14 (12)    \\
\hline
* Renorm./fact. scales, VV         & \NA         & 9          & 4          & \NA          & \NA         & \multirow{8}{*}{2 (1)} \\
* Renorm./fact. scales, ZH     & 3.5        & \NA         & \NA         & \NA          & \NA         &            \\
* Renorm./fact. scales, DM     & 5          & \NA         & \NA         & \NA          & \NA         &            \\
* PDF, WZ background            & \NA         & \NA         & 1.5        & \NA          & \NA         &            \\
* PDF, ZZ background              & \NA         & 1.5        & \NA         & \NA          & \NA         &            \\
* PDF, Higgs boson signal        & 1.5        & \NA         & \NA         & \NA          & \NA         &            \\
* PDF, DM signal                      & 1--2        & \NA         & \NA         & \NA          & \NA         &            \\
\hline
* MC sample size, NRB                  & \NA         & \NA         & \NA         & 5           & \NA         & \multirow{6}{*}{1}  \\
* MC sample size, DY                           & \NA         & \NA         & \NA         & \NA          & 30         &            \\
* MC sample size, \ZZ                          & \NA         & 0.1        & \NA         & \NA          & \NA         &            \\
* MC sample size, \WZ                          & \NA         & \NA         & 2          & \NA          & \NA         &            \\
* MC sample size, ZH          & 1          & \NA         & \NA         & \NA          & \NA         &            \\
* MC sample size, DM          & 3          & \NA         & \NA         & \NA          & \NA         &            \\
\hline
NRB extrapolation to the SR         & \NA         & \NA         & \NA         & 20          & \NA         & $<$1       \\
DY extrapolation to the SR          & \NA         & \NA         & \NA         & \NA          & 100        & $<$1     \\
Lepton efficiency (WZ CR)            & \NA         & \NA         & 3          & \NA          & \NA         & $<$1       \\
Nonprompt bkg. (WZ CR)      & \NA         & \NA         & \NA         & \NA          & 30         & $<$1      \\
\hline
Integrated luminosity                          & \multicolumn{5}{c}{2.5}                                         & $<$1       \\
\hline
* Electron efficiency                          & \multicolumn{5}{c}{1.5}                                         & \multirow{9}{*}{1 ($<$1)}  \\
* Muon efficiency                              & \multicolumn{5}{c}{1}                                           &           \\
* Electron energy scale                        & \multicolumn{5}{c}{1--2}                                         &           \\
* Muon energy scale                            & \multicolumn{5}{c}{1--2}                                         &           \\
* Jet energy scale                             & \multicolumn{5}{c}{1--3 (typically anticorrelated w/ yield)}   &           \\
* Jet energy resolution                        & \multicolumn{5}{c}{1 (typically anticorr.)}                &           \\
* Unclustered energy ($\ptmiss$)            & \multicolumn{5}{c}{1--4 (typically anticorr.), strong in DY}&           \\
* Pileup                                       & \multicolumn{5}{c}{1 (typically anticorrelated)}                &           \\
* b tagging eff. \& mistag rate                                & \multicolumn{5}{c}{1}                                         &           \\
\hline
* BDT: electron energy scale & 1.1 & 2.9 & 2.6 & \NA & \NA & \multirow{3}{*}{ $\NA$ (2) } \\
* BDT: muon energy scale & 1.5 & 4.3 & 2.7 & \NA & \NA \\
* BDT: $\ptmiss$ scale & 1.0 & 3.2 & 4.1 & \NA & \NA \\
\hline
\end{tabular}
}
\label{tab:syst}
\end{table*}

\section{Results}
\label{sec:results}

\begin{figure*}[htbp]
\centering
\includegraphics[width=\cmsSingleFigWidth]{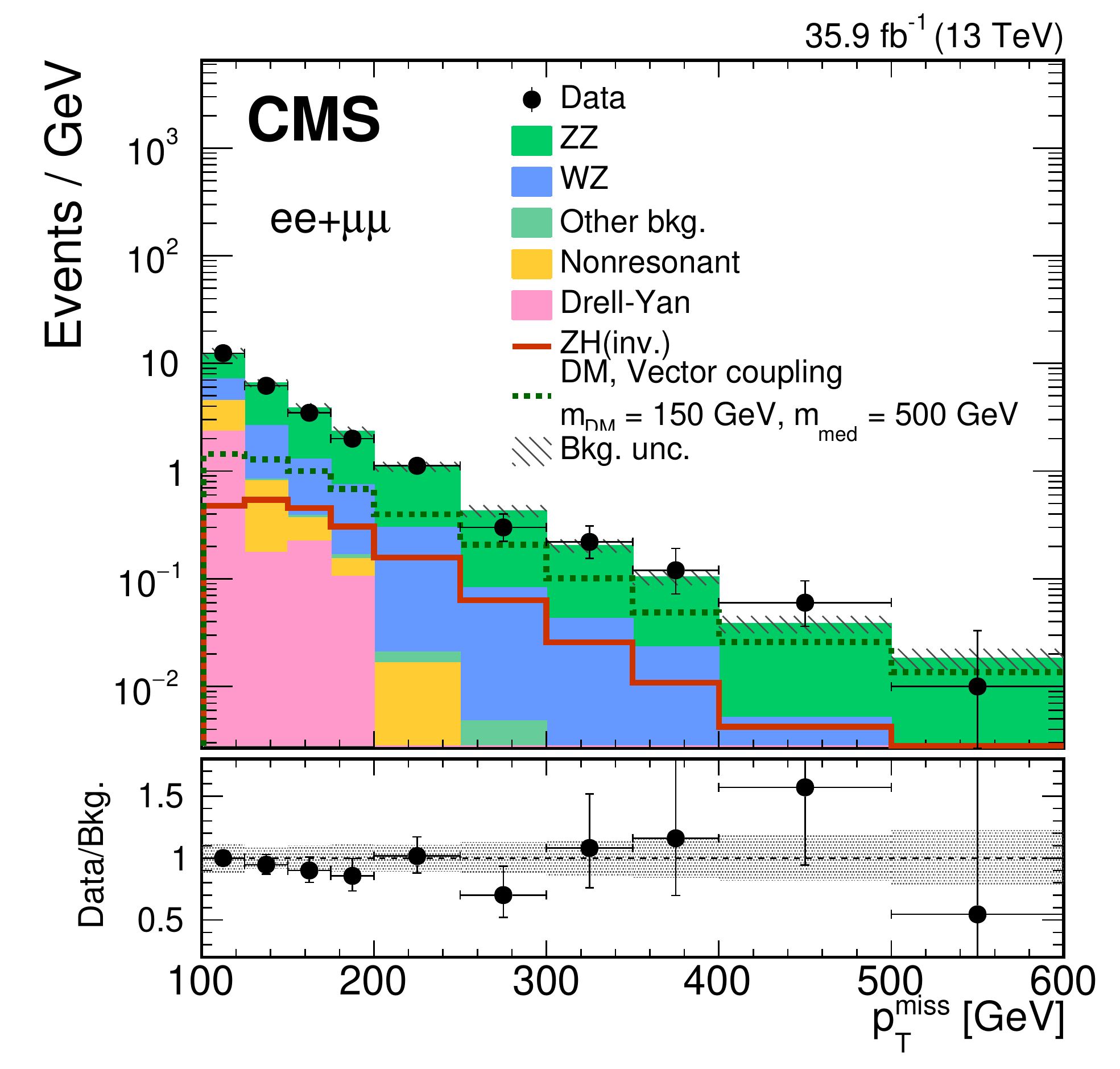}
\caption{
Distribution of the $\ptmiss$ in the combination of the $\Pe\Pe$ and $\PGm\PGm$ channels after the full selection.
The last bin also includes any events with $\ptmiss>600\GeV$.
The uncertainty band includes both statistical and systematic components.
The $\ZHinv$ signal normalization assumes SM production rates and the branching fraction $\mathcal{B}(\Hi \to \text{inv.}) = 1$.
}
\label{fig:distributions}

\end{figure*}

The numbers of observed and expected events for the \ptmiss-based analysis are shown in Table~\ref{tab:zhinvsel}.
There is no significant difference between the dielectron and dimuon channels
in terms of signal-to-background ratio, and hence both are treated together when obtaining the final results.
The observed number of events in the $\Pe\Pe$ ($\PGm\PGm$) channel is 292 (406), and the number of events expected from simulation is $301\pm23$ ($391\pm26$).
Figure~\ref{fig:distributions} shows the $\ptmiss$ distribution in the $\Pe\Pe+\PGm\PGm$ channel in the SR.
The total background estimates and the observed numbers of events in each $\ptmiss$ bin are listed in Table~\ref{tab:binnedCounts_sr},
for both a combined background-only fit to the SR and the CRs, as well as for a fit to the CRs only.
The latter results can be used in conjunction with the SR bin correlation matrix presented in \suppMaterial to recast these results in
the simplified likelihood framework~\cite{simplified-likelihood}.

\begin{table}[hbtp]
\topcaption{
Signal predictions, post-fit background estimates, and observed numbers of events in the \ptmiss-based analysis.
The combined statistical and systematic uncertainties are reported.
\label{tab:zhinvsel}}
\centering
{
\begin{tabular}{lc}
\hline
Process                                           & \multicolumn{1}{c}{$\Pe\Pe + \PGm\PGm$} \\
\hline
qqZH(inv.)                                        & \multirow{2}{*}{$ 158.6  \pm       5.4\x\x $} \\
{ $\ \  m_{\PH}=125 \GeV$, $\mathcal{B}({\PH \to \text{inv.}}) = 1$} \\
ggZH(inv.)                                        &   \multirow{2}{*}{$42.7 \pm       4.9\x $} \\
{ $\ \  m_{\PH}=125 \GeV$, $\mathcal{B}({\PH \to \text{inv.}}) = 1$} \\
DM, vector mediator                               &  \multirow{2}{*}{$ 98.8 \pm        3.9\x $} \\
{ $\ \  m_\text{med}=500 \GeV$, $m_\mathrm{DM} = 150 \GeV$} \\
DM, axial-vector mediator                         & \multirow{2}{*}{$  65.5 \pm        2.6\x  $} \\
{ $\ \  m_\text{med}=500 \GeV$, $m_\mathrm{DM} = 150\GeV$} \\
\hline
ZZ                                                &  $ 379.8 \pm   	 9.4\x\x $  \\
WZ                                                &  $ 162.5 \pm    	 6.8\x\x  $ \\
Nonresonant bkg.                                  &  $ 75   \pm    	15 $  \\
Drell--Yan                                        & $  72   \pm    	29 $  \\
Other bkg.                                        &  $  2.6 \pm    	 0.2  $ \\
\hline
Total bkg.                                        & $  692   \pm       35\x  $ \\
\hline
Data                                              &  698                 \\
\hline
\end{tabular}
}
\end{table}

\begin{table*}[hbtp]
\topcaption{
Expected event yields in each $\ptmiss$ bin for the sum of background processes in the SR.
The background yields and their corresponding uncertainties are obtained after performing a fit to data.
Two sets of background yields are reported: one from a background-only fit to data in both the SR and the CRs, and one from a fit to data in all CRs, but excluding data in the SR.
The observed numbers of events in each bin are also included.
}
\label{tab:binnedCounts_sr}
\centering
{
\begin{tabular}{l c D{,}{\,\pm\,}{-1} D{,}{\,\pm\,}{-1}}
\hline
\multirow{2}{*}{$\ptmiss$ bin (\GeVns{})}     & \multirow{2}{*}{Observed events} & \multicolumn{2}{c}{Total background prediction} \\
&                                     & \multicolumn{1}{c}{SR+CR fit} & \multicolumn{1}{c}{CR-only fit} \\
\hline
$100 \leq \ptmiss < 125$  & 311                                 & 300, 18                       & 256,32           \\
$125 \leq \ptmiss < 150$  & 155                                 & 155.0, 7.0                    & 150,12           \\
$150 \leq \ptmiss < 175$  & 87                                  & 90.8, 4.6                     & 86.9,8.4         \\
$175 \leq \ptmiss < 200$  & 50                                  & 54.7, 3.1                     & 52.7,5.3         \\
$200 \leq \ptmiss < 250$  & 56                                  & 51.3, 2.9                     & 50.2,4.9         \\
$250 \leq \ptmiss < 300$  & 15                                  & 19.7, 1.4                     & 19.4,2.2         \\
$300 \leq \ptmiss < 350$  & 11                                  & 9.64, 0.80                    & 9.4,1.2          \\
$350 \leq \ptmiss < 400$  & 6                                   & 4.73, 0.47                    & 4.58,0.66        \\
$400 \leq \ptmiss < 500$  & 6                                   & 3.44, 0.39                    & 3.31,0.54        \\
\multicolumn{1}{c}{$ \ptmiss  \geq 500 $}           & 1                                   & 1.63, 0.24                    & 1.57,0.33        \\
\hline
\end{tabular}
}

\end{table*}

No deviation from the SM background expectation is found.
Upper limits on the contribution of events from new physics are
computed by using the modified frequentist approach
$CL_s$~\cite{junkcls,Read1} based on asymptotic
formulas~\cite{Cowan:2010js,HiggsCombination}, via a simultaneous maximum likelihood fit to the SR and the CRs.
The expected numbers of background events and signal events, scaled by
a signal strength modifier, are combined in a profile likelihood test
statistic, in which the systematic uncertainties are incorporated as
nuisance parameters.  For the dominant backgrounds in the SR, additional
parameters are introduced to link the background expectations in the SR
to their respective contributions in the CRs discussed in Section~\ref{sec:backgrounds}.
To compute limits in all models, a binned likelihood test statistic is employed, based on the $\ptmiss$ distribution in Fig.~\ref{fig:distributions}
and also on the BDT classifier distribution in the case of invisible decays of the SM Higgs boson.

\subsection{Dark matter interpretation}

\begin{figure*}[!hbtp]
\centering
\includegraphics[width=\cmsDoubleFigWidth]{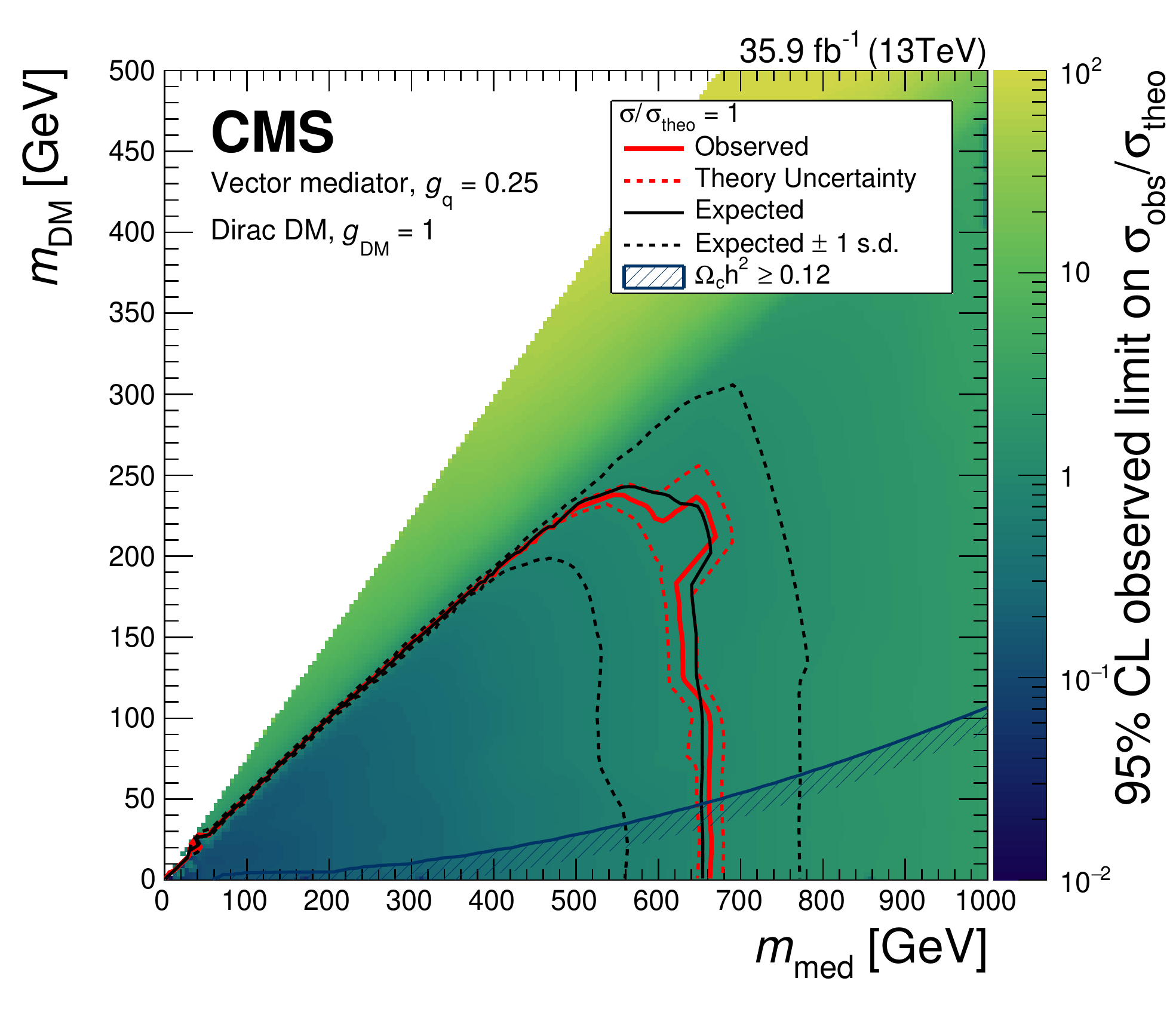} \hfil
\includegraphics[width=\cmsDoubleFigWidth]{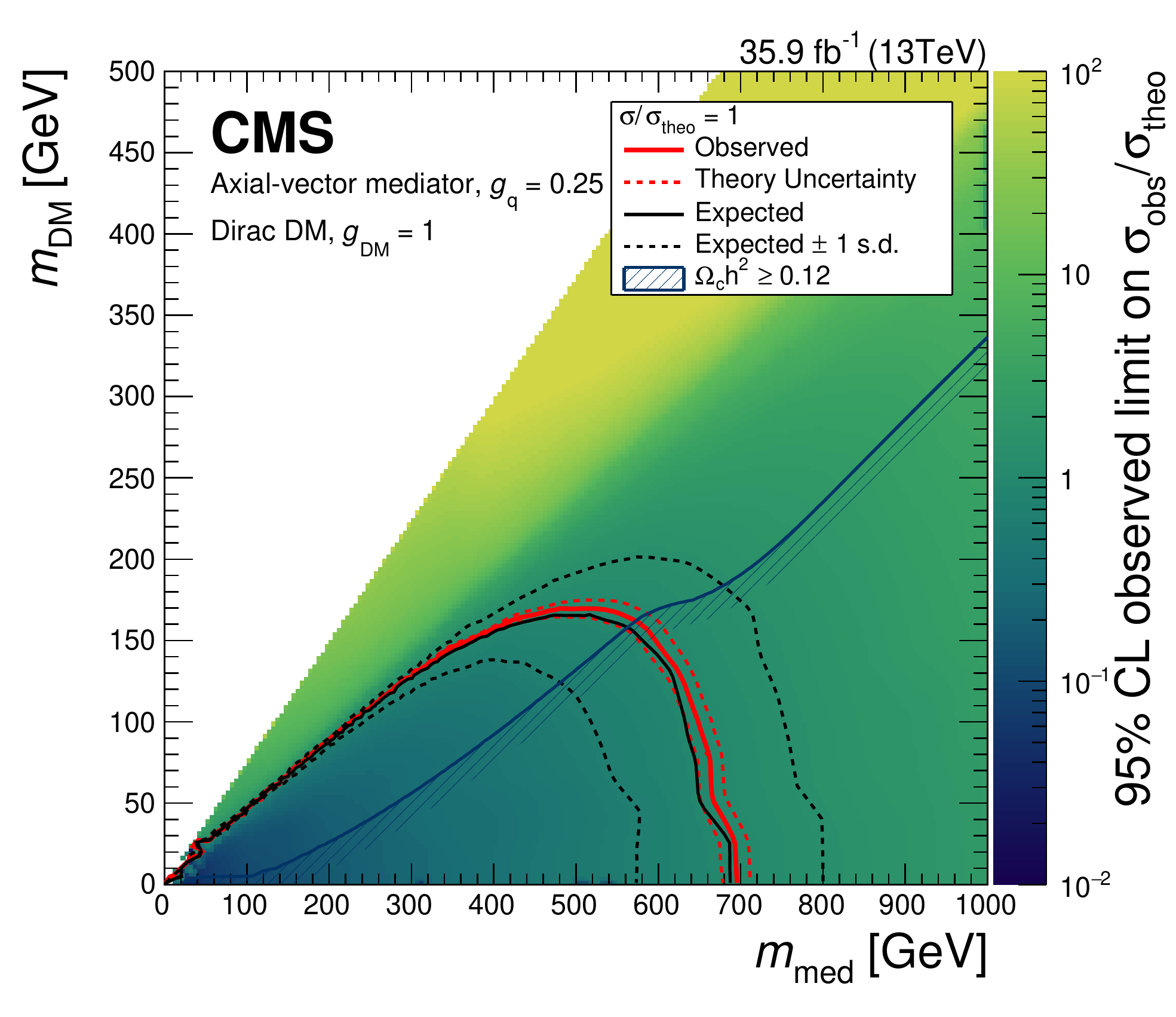}
\caption{
The 95\%~CL expected and observed limits on $\sigma_\text{obs}/\sigma_\text{theo}$
for the vector (left) and axial-vector (right) mediators with $g_{\Pq}=0.25$ and $g_\mathrm{DM} = 1$.
Limits are not shown for far off-shell ($2m_\mathrm{DM} > 1.5 m_\text{med}$) regions of the parameter space.
}
\label{fig:DM13TeV_MV_MX_gq-0p25}
\end{figure*}

\begin{figure*}[!hbtp]
\centering
\includegraphics[width=\cmsDoubleFigWidth]{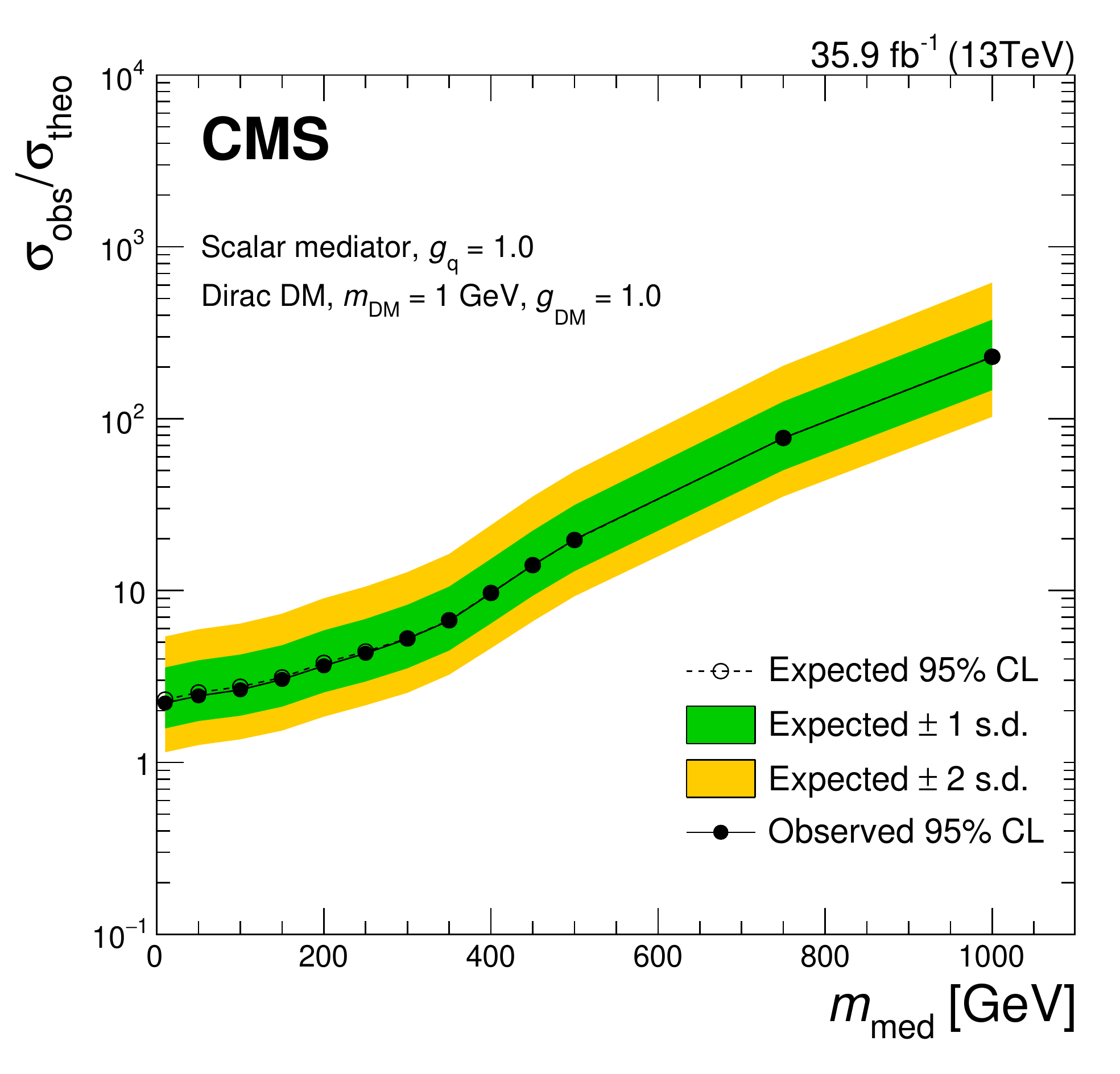} \hfil
\includegraphics[width=\cmsDoubleFigWidth]{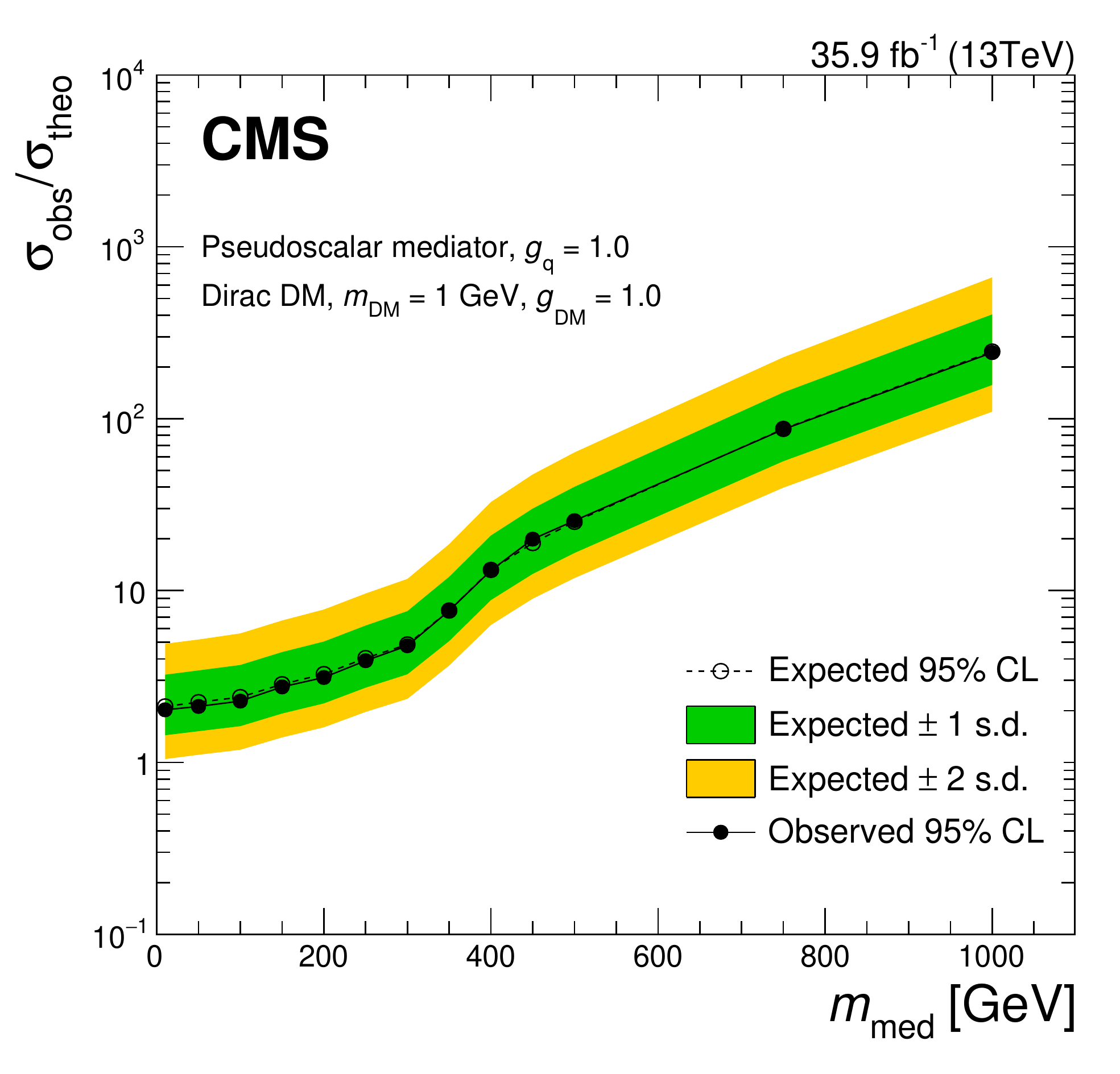}
\caption{
The 95\%~CL expected and observed limits on $\sigma_\text{obs}/\sigma_\text{theo}$
for the scalar (left) and pseudoscalar (right) mediated DM scenario with $g_{\Pq}=g_\mathrm{DM}=1$.
The limits are parameterized as a function of mediator mass $m_\text{med}$ for a fixed dark matter mass $m_\mathrm{DM}=1\GeV$.
}
\label{fig:DM13TeV_MS_MX_gq-1}
\end{figure*}

\begin{figure*}[!hbtp]
\centering
\includegraphics[width=\cmsDoubleFigWidth]{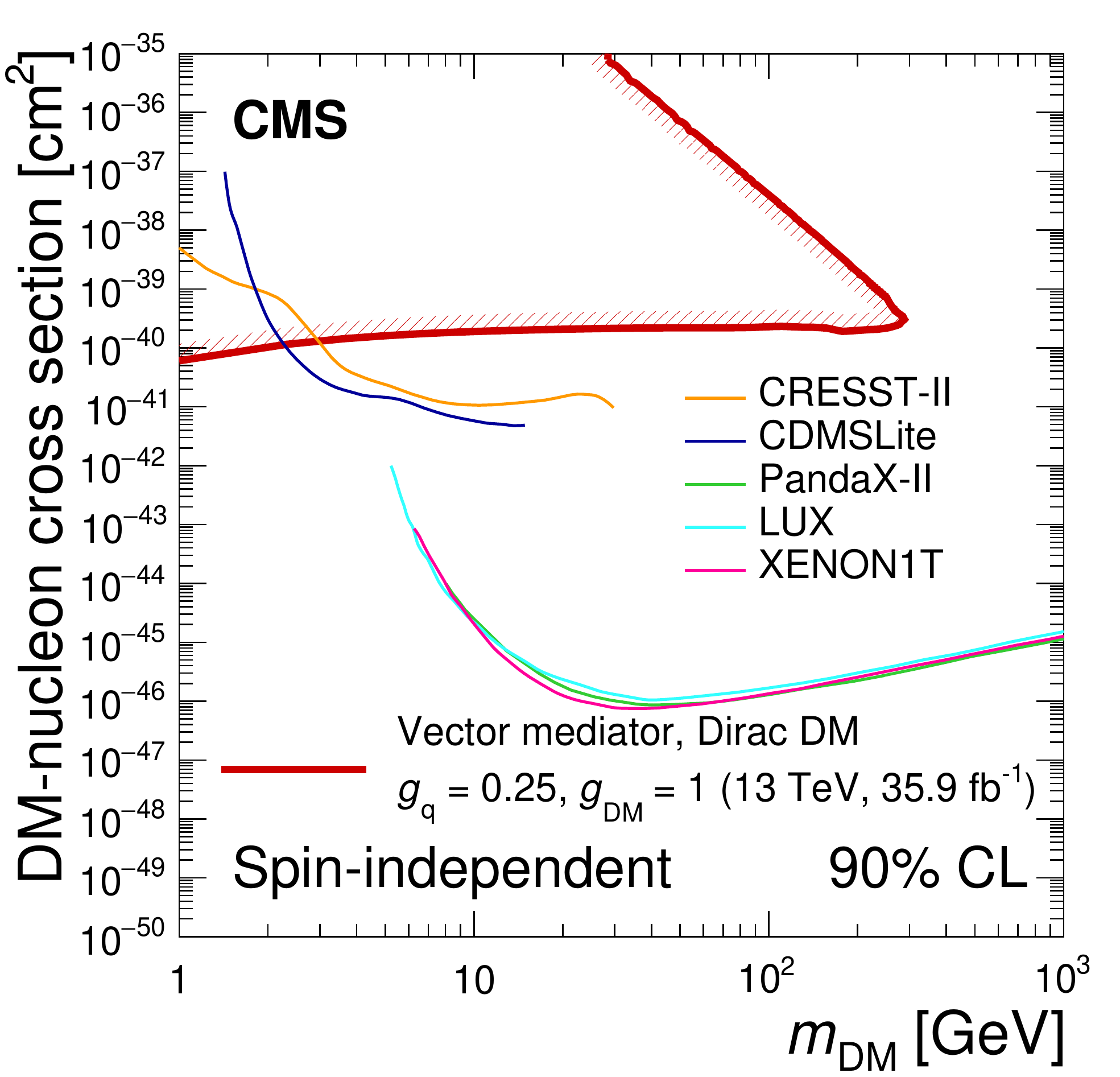} \hfil
\includegraphics[width=\cmsDoubleFigWidth]{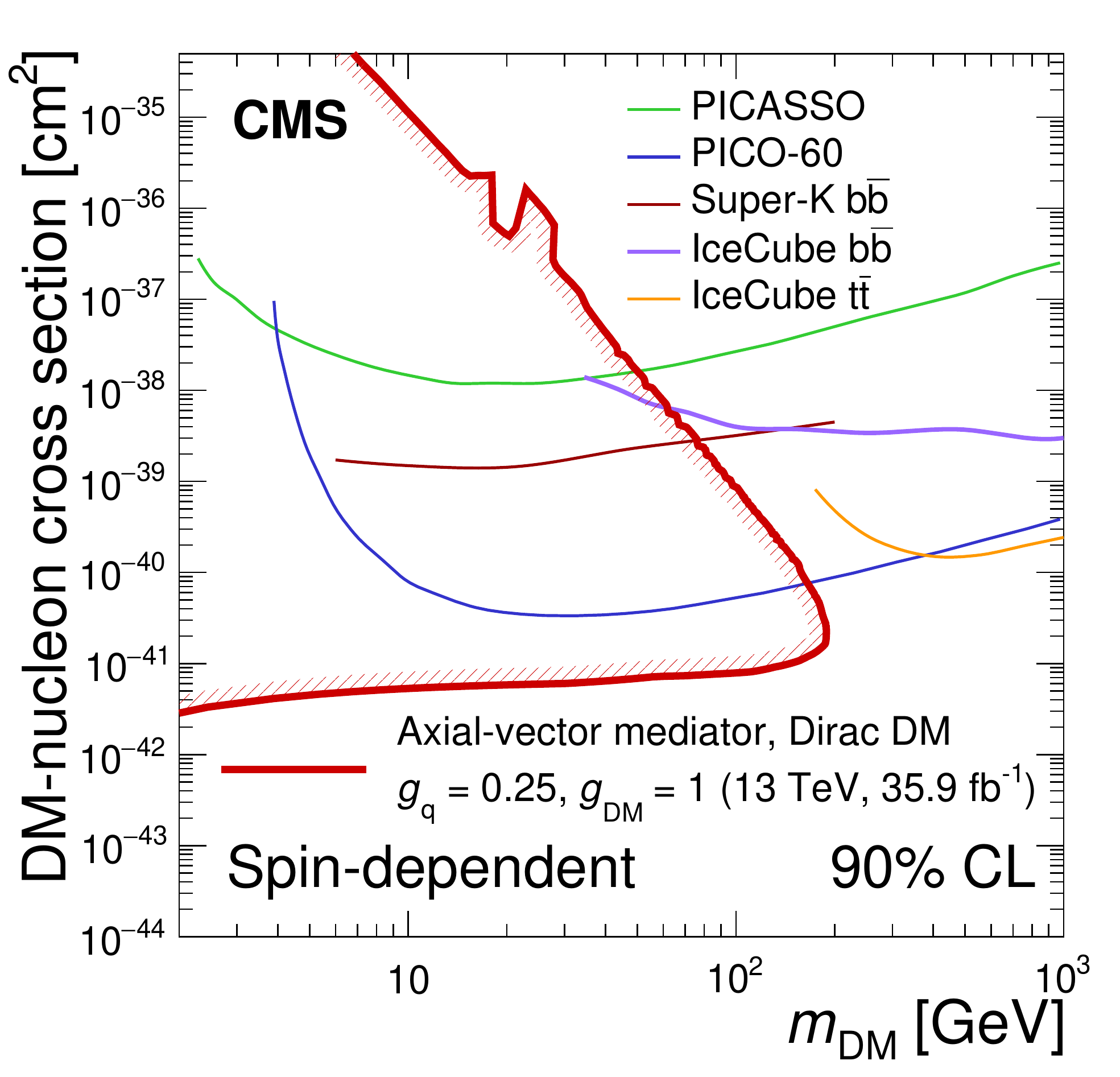}
\caption{Observed 90\% CL limits on the DM-nucleon scattering cross sections
in both spin-independent (left) and spin-dependent (right) cases,
assuming a mediator-quark coupling constant $g_{\Pq} = 0.25$ and mediator-DM coupling constant $g_\mathrm{DM} = 1$.
Limits from the CRESST-II~\cite{Angloher:2015ewa}, \mbox{CDMSLite}~\cite{Agnese:2015nto}, PandaX-II~\cite{Cui:2017nnn}, LUX~\cite{Akerib:2016vxi}, and XENON1T~\cite{Aprile:2017iyp} experiments
are shown for the spin-independent case (vector couplings).
Limits from the PICASSO~\cite{Behnke:2016lsk}, PICO-60~\cite{Amole:2017dex}, Super-Kamiokande~\cite{Choi:2015ara}, and IceCube~\cite{Aartsen:2016zhm,Aartsen:2016exj} experiments
are shown for the spin-dependent case (axial-vector couplings).
}
\label{fig:DDlimits}
\end{figure*}

Figure~\ref{fig:DM13TeV_MV_MX_gq-0p25} shows the 95\%~CL expected and observed limits for
vector and axial-vector scenarios with couplings $g_{\Pq}=0.25$, $g_\mathrm{DM}=1$.
Figure~\ref{fig:DM13TeV_MS_MX_gq-1} shows the 95\%~CL expected and observed limits for
couplings $g_{\Pq}=g_\mathrm{DM}= 1$ in the scalar and pseudoscalar scenarios.
In Fig.~\ref{fig:DDlimits}, limits on the DM-nucleon scattering cross section are set at 90\% CL as a function of the DM particle mass
and compared to selected results from direct detection experiments.
Both spin-dependent and spin-independent cases are considered. In both cases, couplings $g_{\Pq}=0.25$ and
$g_\mathrm{DM}=1$ are used.

\subsection{Limits on invisible Higgs boson decays}

\begin{figure}[htbp]
\centering
\includegraphics[width=\cmsSingleFigWidth]{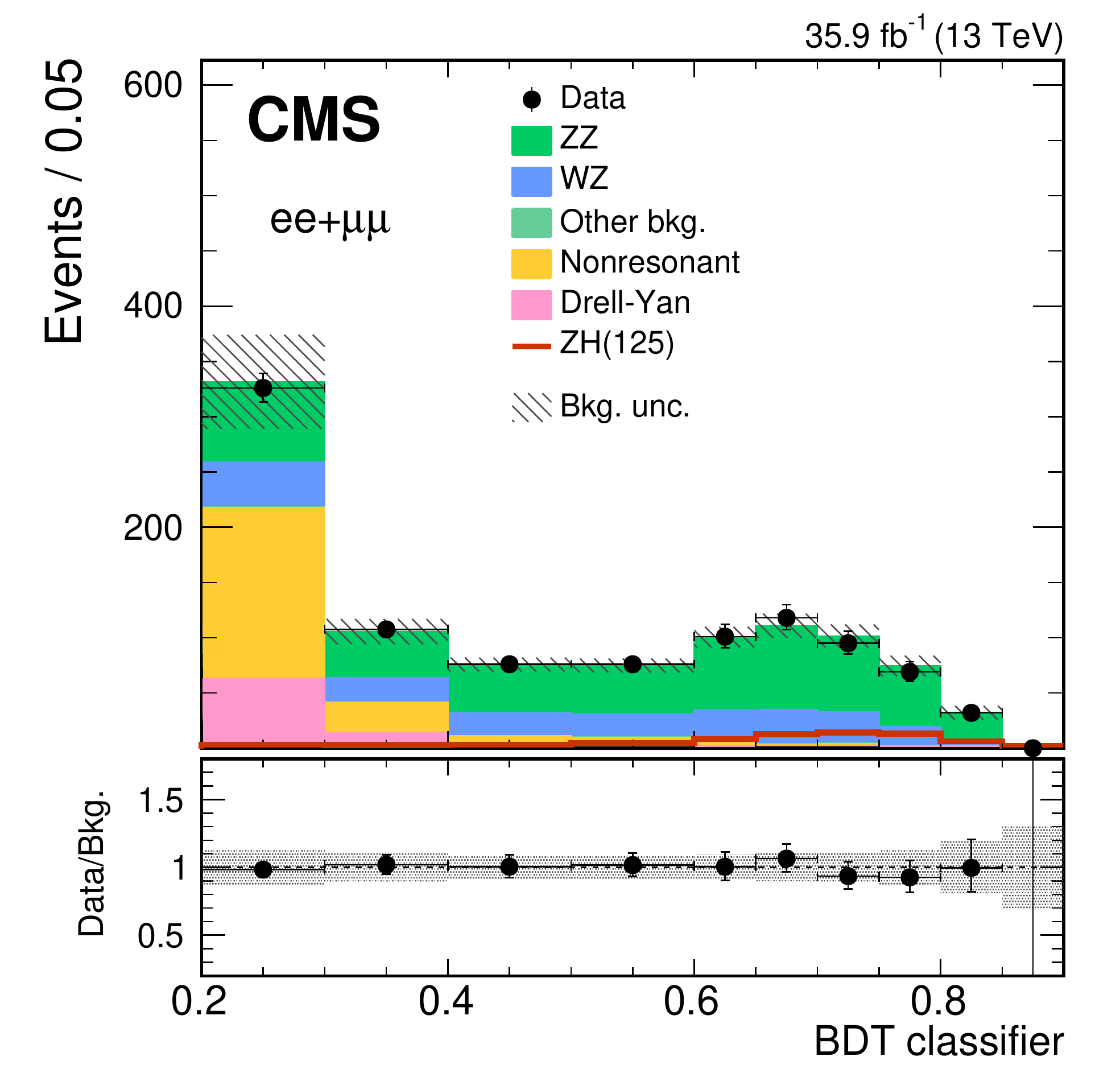}
\caption{Post-fit distribution of the BDT classifier in the multivariate analysis signal region for the SM H(inv.) decay hypothesis with $\mathcal{B}(\PH \to \text{inv.}) = 100\%$. Uncertainty bands correspond to the combined statistical and systematic components.}
\label{fig:bdt_zh}

\end{figure}

\begin{figure}[htbp]
\centering
\includegraphics[width=\cmsSingleFigWidth]{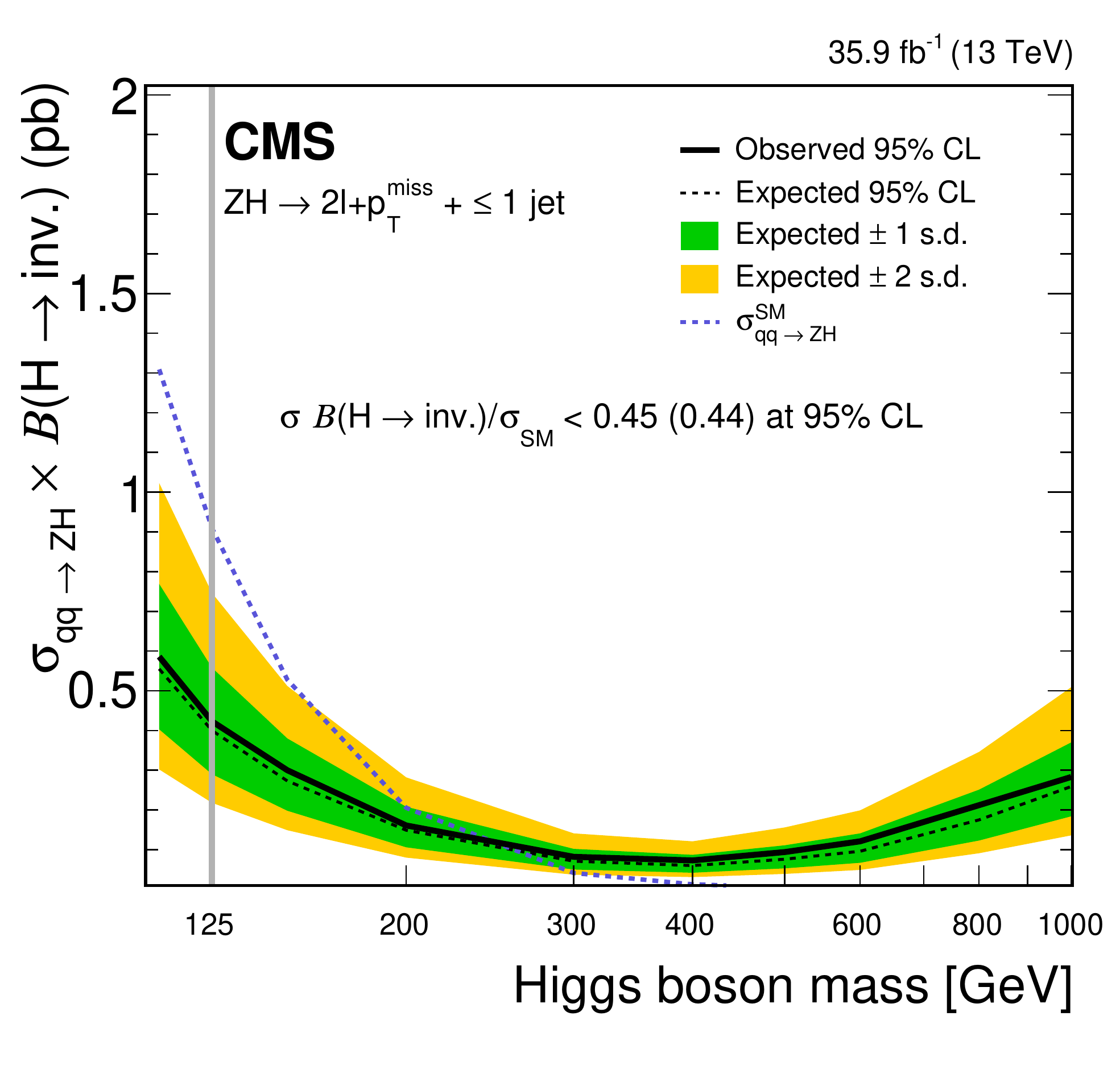}
\caption{Expected and observed 95\% CL upper
limits on the product of the production cross section and the branching fraction,
$\sigma_{{\PQq\PQq} \to \ZH} \, \mathcal{B}(\Hi \to \text{inv.})$, as a function of the SM-like Higgs boson mass.
The limits consider only quark-induced Higgs boson production.
In addition, for the SM (125\GeV) Higgs boson, the limit on branching fraction assuming SM production rate (considering also gluon fusion) is presented.
The vertical gray line indicates that the result at $m_\Hi=125\GeV$ should not be read from the plot, as the gluon contribution is known for that point.
}
\label{fig:xsLim}

\end{figure}

Upper limits are derived for the Higgs boson production cross section
using the same $\ptmiss$-shape analysis as for the DM model.
In addition, for $\mHi = 125\GeV$, a shape analysis using the multivariate classifier distribution, as described in
Section~\ref{sec:mva}, is performed. The resulting post-fit signal region is shown in Fig.~\ref{fig:bdt_zh}.
The 95\% CL expected and observed upper limits
on the product of the production cross section and the branching fraction,
$\sigma_{\ZH} \, \mathcal{B}(\Hi \to \text{inv.})$, computed with the asymptotic $CL_s$
method are shown as a function of the SM-like Higgs boson mass in Fig.~\ref{fig:xsLim} for the $\ptmiss$-shape analysis.
For $\mHi = 125\GeV$, the search can be interpreted as an upper limit on
$\mathcal{B}(\Hi \to \text{inv.})$ assuming the SM production rate of a Higgs boson in association with a $\PZ$ boson.
Assuming the SM production rate, the 95\% observed (expected) CL upper limit on $\mathcal{B}(\Hi \to \text{inv.})$ is
0.45 (0.44) using the $\ptmiss$-shape analysis, and 0.40 (0.42) using the multivariate analysis.
The $\mathrm{gg} \to \Z(\ell\ell)\Hi$ process is considered only for the 125 GeV mass point, and only when interpreting the result as a limit on branching fraction.
For SM-like Higgs production, considering only the $\mathrm{qq}\to \Z(\ell\ell)\Hi$ process, upper limits on $\mathcal{B}(\Hi \to \text{inv.})$ are presented as a function of $\mHi$ in \suppMaterial.

\subsection{Unparticle interpretation}

\begin{figure}[!hbtp]
\centering
\includegraphics[width=\cmsSingleFigWidth]{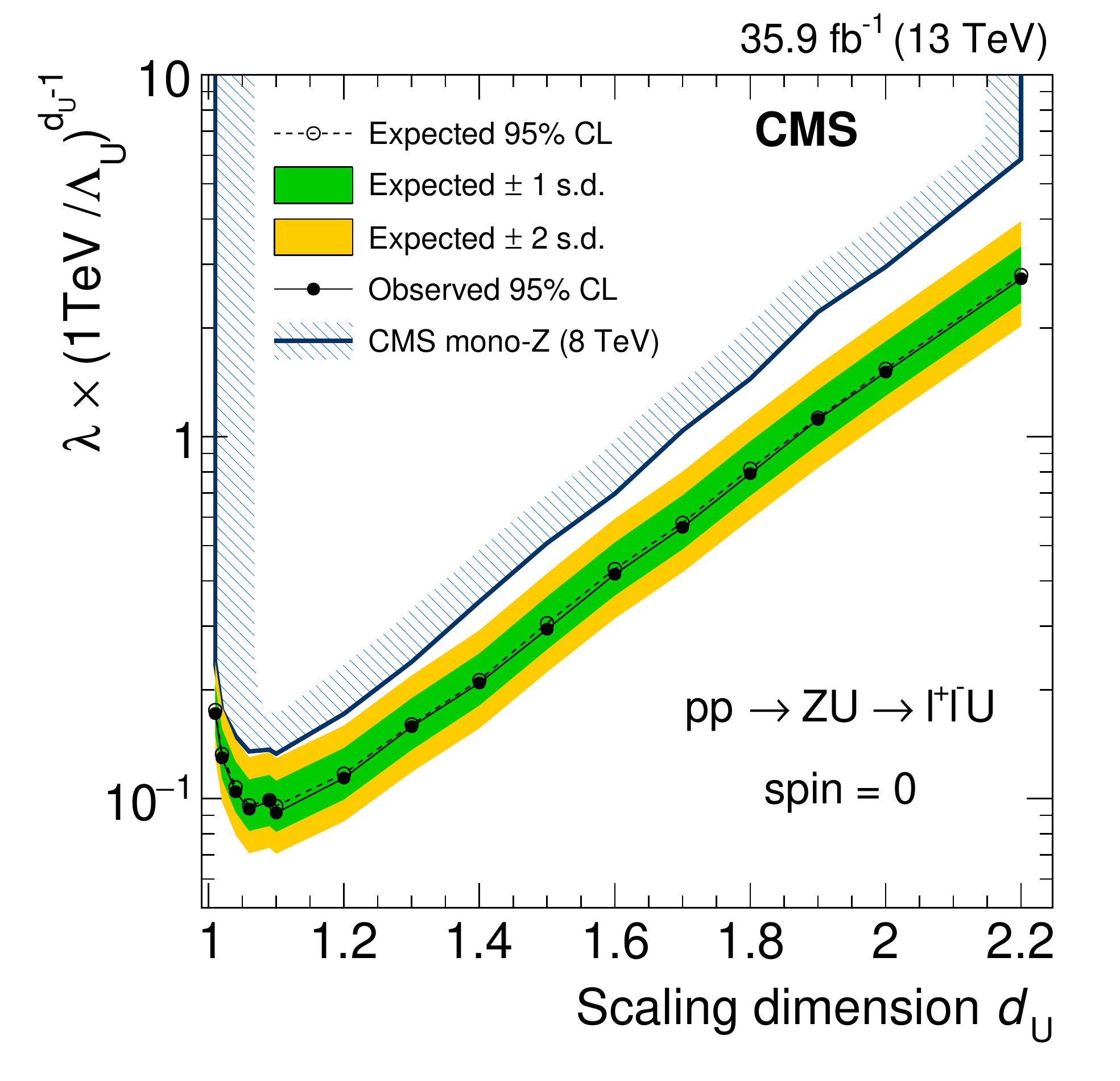}
\caption{
The 95\% CL upper limits on the Wilson coefficient $\lambda / \LU^{\dU-1}$ of the unparticle-quark coupling operator.
The results from an earlier CMS search in the same final state~\cite{Khachatryan:2015bbl} are shown for comparison.
}
\label{fig:unparticleLimits}
\end{figure}

In the unparticle scenario, a shape analysis of the $\ptmiss$ spectrum is performed.
Upper limits are set at 95\% CL on the Wilson coefficient $\lambda / \LU^{\dU-1}$ of the unparticle-quark coupling operator,
and are shown in Fig.~\ref{fig:unparticleLimits} as a function of the scaling dimension $d_\textsf{U}$.

\subsection{The ADD interpretation}
\begin{figure*}[!hbtp]
\centering
\includegraphics[width=\cmsDoubleFigWidth]{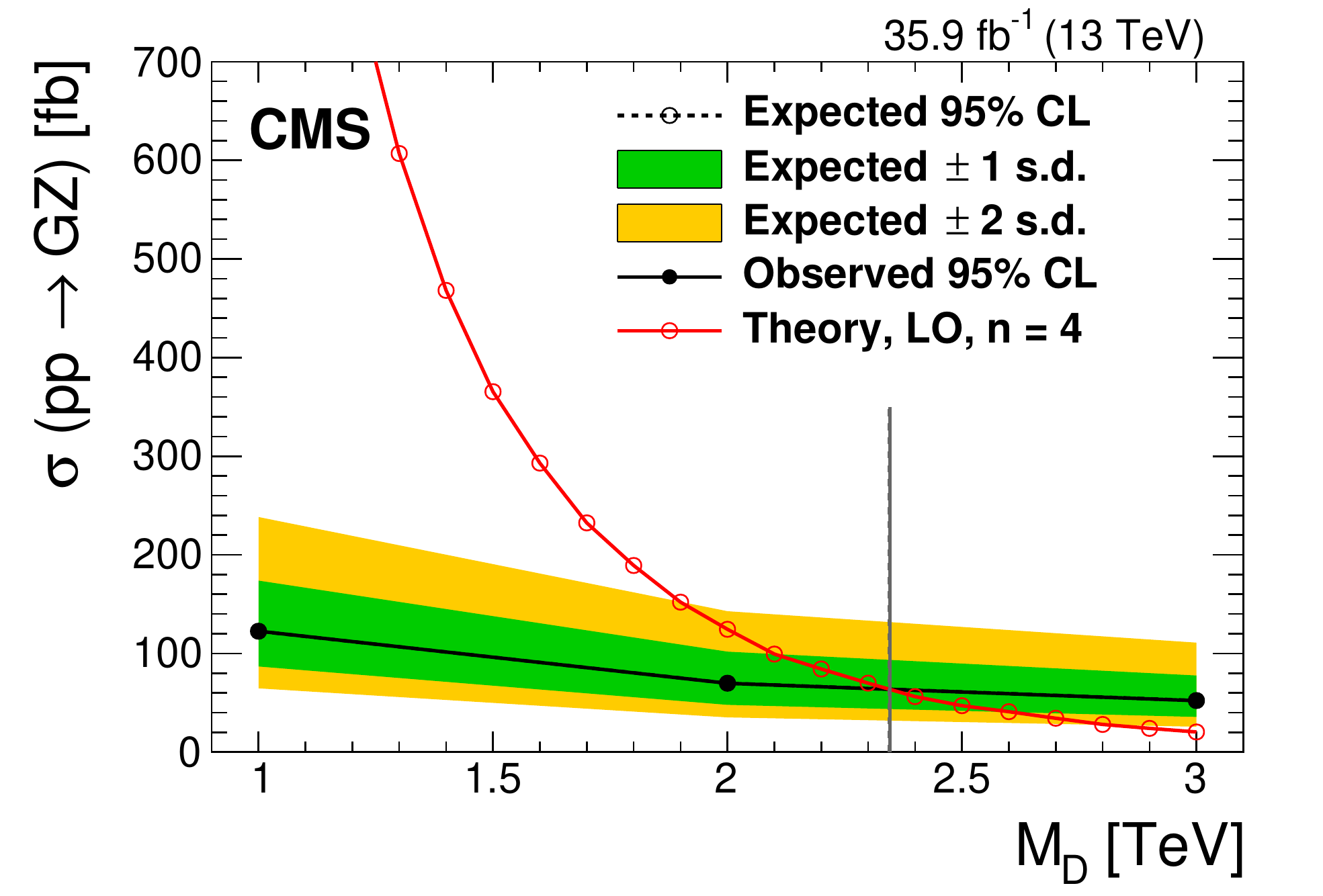} \hfil
\includegraphics[width=\cmsDoubleFigWidth]{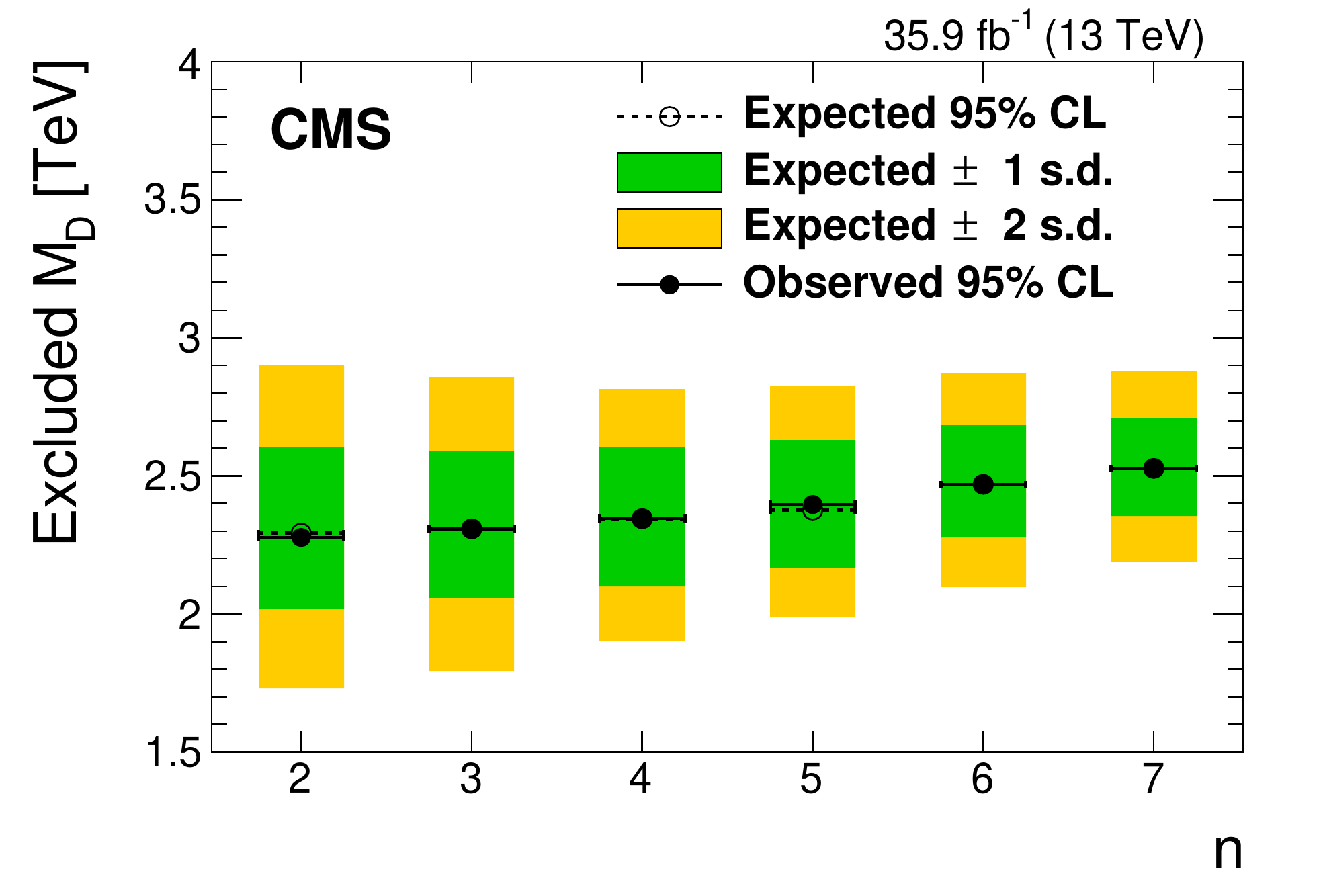}
\caption{Expected and observed 95\% CL cross section exclusion limits for the example case $n=4$ in the ADD scenario (left) and exclusion limits on $M_\mathrm{D}$ for different values of $n$ (right).
In both plots, the markers for the expected exclusion are obscured by the close overlap with those for the observed exclusion.
The red solid line in the left plot shows the theoretical cross section for the case $n = 4$. Cross sections are calculated in the fiducial phase space of $\pt^{\rm G} > 50\GeV$.
The vertical line in the left plot shows the projection onto the $M_\mathrm{D}$ axis of the intersection of the theory curve with the expected and observed exclusion limits.
}
\label{fig:limits_add}
\end{figure*}

In the framework of the ADD model of large extra dimensions, we calculate limits depending on the number of extra dimensions $n$ and the fundamental Planck scale $M_\mathrm{D}$.
For each value of $n$, cross section limits are calculated as a function of $M_\mathrm{D}$.
By finding the intersection between the theory cross section line, calculated in the fiducial phase space of the graviton transverse momentum $\pt^{\rm G} > 50\GeV$,
with the observed and expected excluded cross sections, and projecting that point onto the $M_\mathrm{D}$ axis,
we find limits on $M_\mathrm{D}$ as a function of $n$, as shown in Fig.~\ref{fig:limits_add}.

The observed and expected exclusion of $M_\mathrm{D}$ ranges between 2.3 and 2.5\TeV for $n$ between 2 and 7, at 95\% CL.

\section{Summary}
\label{sec:summary}

A search for new physics in events with a leptonically decaying $\PZ$ boson and large missing transverse momentum has been presented.
The search is based on a data set of proton-proton collisions collected with the CMS experiment in 2016, corresponding to an integrated luminosity of $\usedLumiWithSyst$ at $\sqrt{s} = 13\TeV$.
No evidence for physics beyond the standard model is found.
Compared to the previous search in the same final state~\cite{CMS-PAPER-EXO-16-010}, the exclusion limits on dark matter and mediator masses are significantly extended for spin-1 mediators in the simplified model interpretation, and exclusion limits for unparticles are also extended.
Results for dark matter production via spin-0 mediators in the simplified model interpretation, as well as graviton emission in a model with large extra dimensions, are presented in this final state for the first time.
In the case of invisible decays of a standard-model-like Higgs boson, the upper limit of 40\% on their branching fraction is set at 95\% confidence level, using data not included in the previously published combined analysis~\cite{CMS-PAPER-HIG-16-016}.

\ifthenelse{\boolean{cms@external}}{}{\clearpage}

\begin{acknowledgments}
We congratulate our colleagues in the CERN accelerator departments for the excellent performance of the LHC and thank the technical and administrative staffs at CERN and at other CMS institutes for their contributions to the success of the CMS effort. In addition, we gratefully acknowledge the computing centres and personnel of the Worldwide LHC Computing Grid for delivering so effectively the computing infrastructure essential to our analyses. Finally, we acknowledge the enduring support for the construction and operation of the LHC and the CMS detector provided by the following funding agencies: BMWFW and FWF (Austria); FNRS and FWO (Belgium); CNPq, CAPES, FAPERJ, and FAPESP (Brazil); MES (Bulgaria); CERN; CAS, MoST, and NSFC (China); COLCIENCIAS (Colombia); MSES and CSF (Croatia); RPF (Cyprus); SENESCYT (Ecuador); MoER, ERC IUT, and ERDF (Estonia); Academy of Finland, MEC, and HIP (Finland); CEA and CNRS/IN2P3 (France); BMBF, DFG, and HGF (Germany); GSRT (Greece); OTKA and NIH (Hungary); DAE and DST (India); IPM (Iran); SFI (Ireland); INFN (Italy); MSIP and NRF (Republic of Korea); LAS (Lithuania); MOE and UM (Malaysia); BUAP, CINVESTAV, CONACYT, LNS, SEP, and UASLP-FAI (Mexico); MBIE (New Zealand); PAEC (Pakistan); MSHE and NSC (Poland); FCT (Portugal); JINR (Dubna); MON, RosAtom, RAS, RFBR and RAEP (Russia); MESTD (Serbia); SEIDI, CPAN, PCTI and FEDER (Spain); Swiss Funding Agencies (Switzerland); MST (Taipei); ThEPCenter, IPST, STAR, and NSTDA (Thailand); TUBITAK and TAEK (Turkey); NASU and SFFR (Ukraine); STFC (United Kingdom); DOE and NSF (USA).

\hyphenation{Rachada-pisek} Individuals have received support from the Marie-Curie programme and the European Research Council and Horizon 2020 Grant, contract No. 675440 (European Union); the Leventis Foundation; the A. P. Sloan Foundation; the Alexander von Humboldt Foundation; the Belgian Federal Science Policy Office; the Fonds pour la Formation \`a la Recherche dans l'Industrie et dans l'Agriculture (FRIA-Belgium); the Agentschap voor Innovatie door Wetenschap en Technologie (IWT-Belgium); the Ministry of Education, Youth and Sports (MEYS) of the Czech Republic; the Council of Science and Industrial Research, India; the HOMING PLUS programme of the Foundation for Polish Science, cofinanced from European Union, Regional Development Fund, the Mobility Plus programme of the Ministry of Science and Higher Education, the National Science Center (Poland), contracts Harmonia 2014/14/M/ST2/00428, Opus 2014/13/B/ST2/02543, 2014/15/B/ST2/03998, and 2015/19/B/ST2/02861, Sonata-bis 2012/07/E/ST2/01406; the National Priorities Research Program by Qatar National Research Fund; the Programa Severo Ochoa del Principado de Asturias; the Thalis and Aristeia programmes cofinanced by EU-ESF and the Greek NSRF; the Rachadapisek Sompot Fund for Postdoctoral Fellowship, Chulalongkorn University and the Chulalongkorn Academic into Its 2nd Century Project Advancement Project (Thailand); the Welch Foundation, contract C-1845; and the Weston Havens Foundation (USA).

\end{acknowledgments}

\bibliography{auto_generated}

\ifthenelse{\boolean{cms@external}}{}{
\appendix
\numberwithin{table}{section}
\numberwithin{figure}{section}
\section{Supplementary Material\label{app:suppMat}}
\input{supplemental_material}
}
\cleardoublepage \section{The CMS Collaboration \label{app:collab}}\begin{sloppypar}\hyphenpenalty=5000\widowpenalty=500\clubpenalty=5000\input{EXO-16-052-authorlist.tex}\end{sloppypar}
\end{document}

%% file: supplemental_material.tex
\begin{figure}[hbtp]
\centering
\includegraphics[width=\cmsSingleFigWidth]{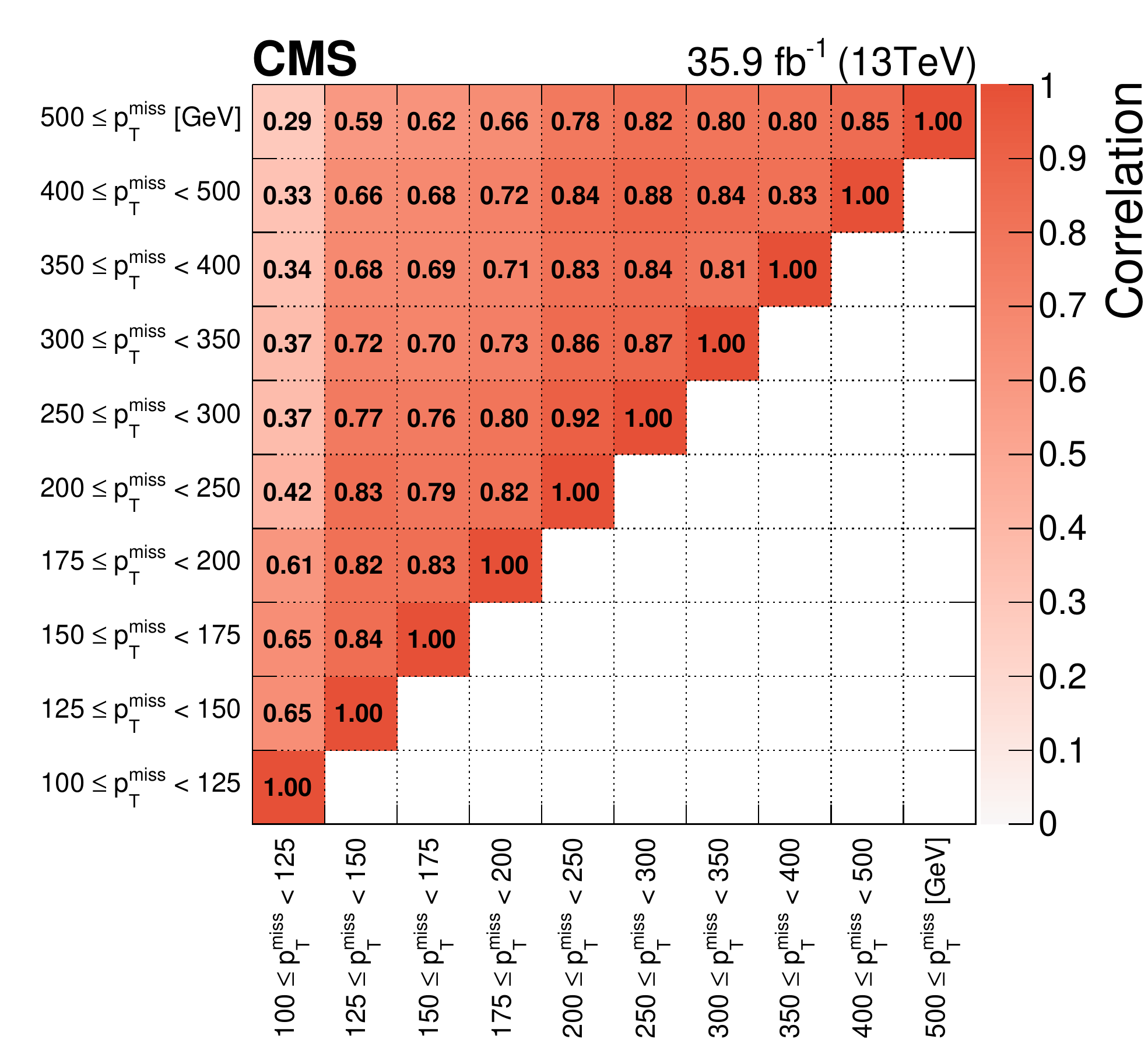}
\caption{
Correlations between the estimated background yields in the signal region $\ptmiss$ bins.
The correlations are obtained after performing a combined fit to data in all control regions, but excluding data in the signal region. Since the correlation matrix is symmetric by construction, the part below the diagonal is not shown.
}
\label{fig:correlation}
\end{figure}

Figure~\ref{fig:correlation} shows the correlations between the estimated background yields in $\ptmiss$ bins in the signal region.
These correlations can be used in conjunction with the total background estimates in the signal region from the control-region-only fit
to re-construct an approximate likelihood function for this analysis.
Using this likelihood function, together with an alternative signal model, a fit can be performed to the observed data to recast the results in
the simplified likelihood framework~\cite{simplified-likelihood}.
To utilize the simplified likelihod method, a prediction of the reconstructed event yields in each $\ptmiss$ bin is required.
This is best obtained by using a detector simulation program such as \textsc{delphes}, however a reasonable prediction can be obtained by:
applying a generator-level selection that parallels the reconstruction-level selection described in Section 7, omitting tau lepton and b jet vetoes;
smearing the $\ptmiss$ with a Gaussian kernel of 24\GeV width; and scaling by a reconstruction efficiency of 0.70.

\begin{figure}[h]
\centering
\includegraphics[width=\cmsSingleFigWidth]{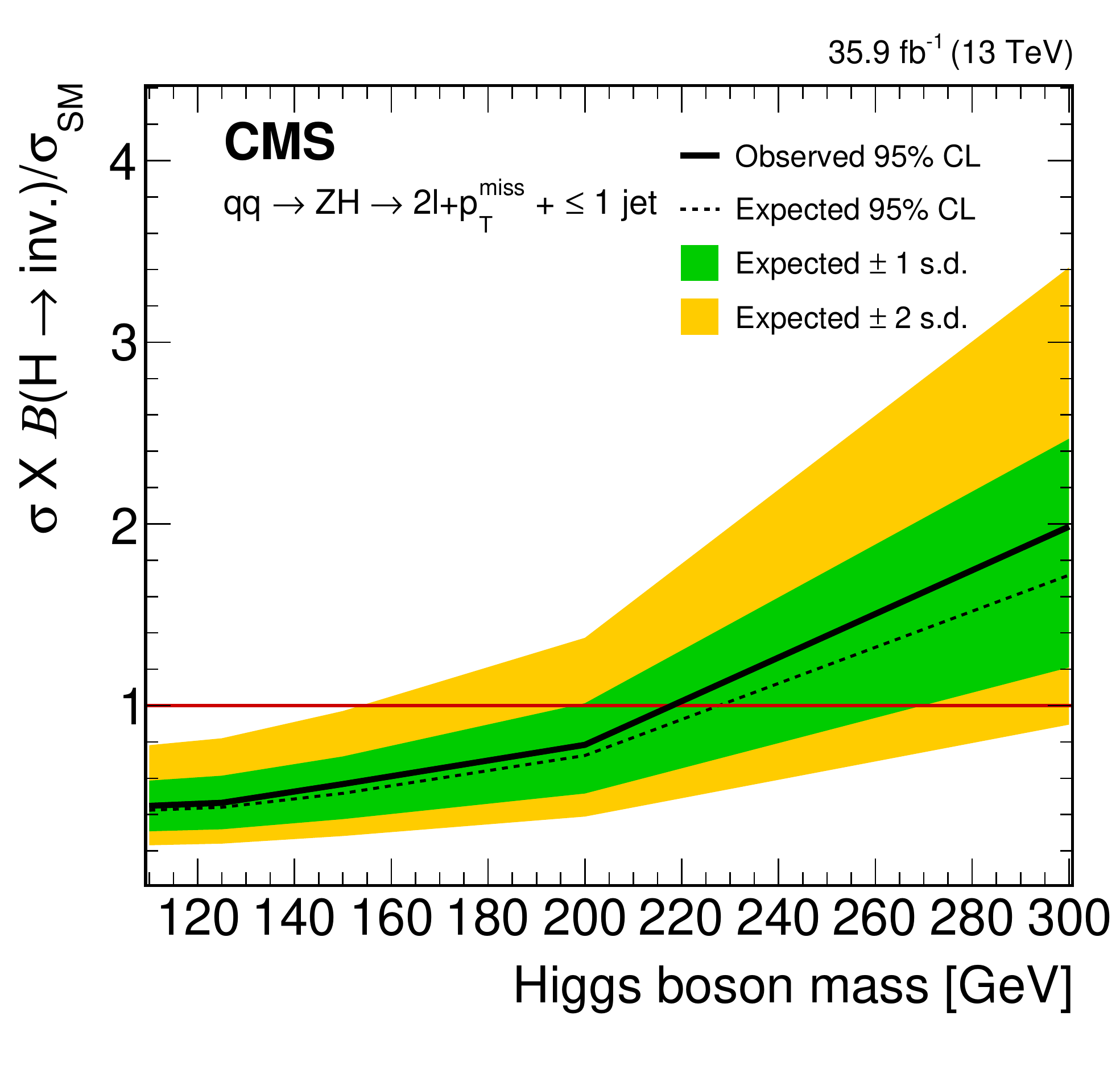}
\caption{
Expected and observed 95\% CL upper limits on $\mathcal{B}(\Hi \to \text{inv.})$, assuming SM Higgs boson production cross sections, as a function of the Higgs boson mass.
}
\label{fig:brLimitScan}
\end{figure}

Figure~\ref{fig:brLimitScan} shows the 95\% CL upper limits on the SM-like Higgs boson branching fraction to invisible particles, as a function of its mass.
The Higgs production cross section assumed in this figure includes only the $\PQq\PAQq \to \PZ(\ell\ell)\Hi$ process.

%% file: EXO-16-052-authorlist.tex
\textbf{Yerevan Physics Institute,  Yerevan,  Armenia}\\*[0pt]
A.M.~Sirunyan, A.~Tumasyan
\vskip\cmsinstskip
\textbf{Institut f\"{u}r Hochenergiephysik,  Wien,  Austria}\\*[0pt]
W.~Adam, F.~Ambrogi, E.~Asilar, T.~Bergauer, J.~Brandstetter, E.~Brondolin, M.~Dragicevic, J.~Er\"{o}, A.~Escalante Del Valle, M.~Flechl, M.~Friedl, R.~Fr\"{u}hwirth\cmsAuthorMark{1}, V.M.~Ghete, J.~Grossmann, J.~Hrubec, M.~Jeitler\cmsAuthorMark{1}, A.~K\"{o}nig, N.~Krammer, I.~Kr\"{a}tschmer, D.~Liko, T.~Madlener, I.~Mikulec, E.~Pree, N.~Rad, H.~Rohringer, J.~Schieck\cmsAuthorMark{1}, R.~Sch\"{o}fbeck, M.~Spanring, D.~Spitzbart, W.~Waltenberger, J.~Wittmann, C.-E.~Wulz\cmsAuthorMark{1}, M.~Zarucki
% ...
\vskip\cmsinstskip
% ...
1:~~Also at Vienna University of Technology, Vienna, Austria\\
% ...
